\documentclass[pra, aps, superscriptaddress, amsmath, amssymb, twocolumn]{revtex4-2}
\usepackage{graphicx, bm, xcolor, dcolumn}
\usepackage[normalem]{ulem}
\usepackage{times}
\usepackage{array}
\usepackage{xcolor}
\usepackage{ulem}
\usepackage{booktabs} 
\usepackage{caption}  
\usepackage{textcomp} 
\usepackage{makecell}

\usepackage[english]{babel}
\usepackage{caption}
\usepackage{microtype} 

\begin{document}

\title{Nonadiabatic effect in high-order harmonic generation revealed by a fully analytical method}

\author{Fengjian Sun}
\affiliation{State Key Laboratory of Ultrafast Optical Science and Technology,  Xi'an Institute of Optics and
Precision Mechanics,  Chinese Academy of Sciences,  Xi'an 710119,  Shaanxi,  China}
\affiliation{Center for Attosecond Science and Technology,  Xi'an Institute of Optics and
Precision Mechanics,  Chinese Academy of Sciences,  Xi'an 710119,  Shaanxi,  China}
\affiliation{Key Laboratory for Physical Electronics and Devices of the Ministry of Education \& Shaanxi Key Lab of Information Photonic Technique,  Xi'an JiaoTong University,  Xi'an 710049,  China}
\affiliation{University of Chinese Academy of Science,  Beijing 100049,  China}

\author{Pei Huang}
\email{huangpei@opt.ac.cn}
\affiliation{State Key Laboratory of Ultrafast Optical Science and Technology,  Xi'an Institute of Optics and
Precision Mechanics,  Chinese Academy of Sciences,  Xi'an 710119,  Shaanxi,  China}
\affiliation{Center for Attosecond Science and Technology,  Xi'an Institute of Optics and
Precision Mechanics,  Chinese Academy of Sciences,  Xi'an 710119,  Shaanxi,  China}

\author{Alexandra S. Landsman}
\affiliation{Department of Physics,  the Ohio State University,  Columbus,  OH 43210,  USA}

\author{Yanpeng Zhang}
\affiliation{Key Laboratory for Physical Electronics and Devices of the Ministry of Education \& Shaanxi Key Lab of Information Photonic Technique,  Xi'an JiaoTong University,  Xi'an 710049,  China}

\author{Liang-Wen Pi}
\email{lwpi@opt.ac.cn}
\affiliation{State Key Laboratory of Ultrafast Optical Science and Technology,  Xi'an Institute of Optics and
Precision Mechanics,  Chinese Academy of Sciences,  Xi'an 710119,  Shaanxi,  China}
\affiliation{Center for Attosecond Science and Technology,  Xi'an Institute of Optics and
Precision Mechanics,  Chinese Academy of Sciences,  Xi'an 710119,  Shaanxi,  China}

\author{Yuxi Fu}
\email{fuyuxi@opt.ac.cn}
\affiliation{State Key Laboratory of Ultrafast Optical Science and Technology,  Xi'an Institute of Optics and
Precision Mechanics,  Chinese Academy of Sciences,  Xi'an 710119,  Shaanxi,  China}
\affiliation{Center for Attosecond Science and Technology,  Xi'an Institute of Optics and
Precision Mechanics,  Chinese Academy of Sciences,  Xi'an 710119,  Shaanxi,  China}

\begin{abstract}


We propose a fully analytical method for describing high-order harmonic generation (HHG). This method is based on the strong-field approximation (SFA) and electron-trajectory theory, but utilizes the perturbation expansion on the Keldysh parameter $\gamma$. This expansion allows us to clearly differentiate the nonadiabatic and adiabatic effects on HHG. We show that the nonadiabatic effect relating to high-order expansion depends on the laser wavelength and remarkably enhances the HHG yields for cases of short wavelengths, providing deeper insights into wavelength-dependent HHG yields which are important in producing attosecond pulses.  Especially, our method provides the analytical and accurate descriptions of nonadiabatic exit velocity and position of the tunneling electron at the tunnel exit. These descriptions are meaningful for constructing a fully analytical and quantitative Coulomb-included HHG model, which is crucial in HHG-based attosecond measurement.

\end{abstract}

\maketitle

\section{Introduction\label{intro}}

When a laser interacts with gases, complex high order nonlinear effects occur,  leading to the HHG. Since the 1980s,  extreme ultraviolet (EUV) HHG have been observed when rare gases are driven by near-infrared intense laser fields {\cite{McPherson:87, Ferray_1988, PhysRevA.39.5751}}. HHG has attracted a lot of attention in the field of optics. In recent decades,  HHG from solids {\cite{Ghimire2019, doi:10.1126/sciadv.aao5207, PhysRevLett.130.166903, PhysRevLett.129.167402, doi:10.34133/2022/9767251}} and liquids {\cite{Luu2018, PhysRevLett.124.203901, PhysRevA.87.063811, Mondal2023}} have also been extensively investigated. As an important method for producing coherent EUV light,  HHG is predicted to have the ability to generate attosecond pulses {\cite{FARKAS1992447}},  which have been observed in experiments {\cite{doi:10.1126/science.1059413, Hentschel2001}}. Additionally,  HHG can be used to probe electron dynamics and molecular structure {\cite{PhysRevA.78.053414, 10.1063/1.3069511,doi:10.34133/ultrafastscience.0034}}. Furthermore,  high photon energy HHG also have the potential to serve as a light source for extreme ultraviolet lithography machines {\cite{doi:10.1126/science.aac9755}}.


Commonly,  ionization is the first step of HHG. It determines the initial conditions of electron dynamics out of the barrier,  and influences HHG yields significantly. The ionization process is usually characterized by the Keldysh parameter {\cite{Keldysh65}},  also known as the adiabatic parameter {\cite{Perelomov1966IONIZATIONOA}},  which can be expressed as $\gamma\equiv \omega \sqrt{2I_p}/ E_0$, where $\omega$,  $E_0$,  and $I_p$ represent the laser frequency (also photon energy in atomic units),  peak electric field strength, and binding energy of the target atom, respectively. When $\gamma\ll 1$,  ionization can be treated as an adiabatic case,  and this process can be described in adiabatic approximation. In the range of intermediate $\gamma\sim  1$,  which is typical for many current intense field experiments, this process will be greatly influenced by the nonadiabatic effect. In 1966, the PPT theory was proposed to evaluate the ionization yield {\cite{Perelomov1966IONIZATIONOA}}, which takes the nonadiabatic effect into account. In 1986, the ADK theory with a compact form was proposed, which works well especially in the adiabatic region for $\gamma\ll 1$ and has been widely adopted {\cite{1986Tunnel}}.

Following these pioneering theoretical works,  the adiabatic approximation and nonadiabatic effects in the ionization process have garnered a lot of attention. Many studies were subsequently carried out to investigate the ionization process in comparison to experimental results. In  {\cite{Delone:91}},  a calculation based on the adiabatic limit condition was presented,  showing good agreement with experimental results for Kr in a linearly polarized laser field {\cite{PhysRevLett.63.2212}}. Additionally,  it was found that the tunnel exit momentum is zero in the adiabatic limit but nonzero in the nonadiabatic region. This characteristic has been widely adopted in theoretical and experimental analyses {\cite{PhysRevLett.105.133002, PhysRevLett.116.063003, PhysRevLett.127.273201, 10.3389fphy.2023.1120654, PhysRevLett.109.083002}},  making it a useful tool in distinguishing between adiabatic and nonadiabatic methods. Under the adiabatic approximation,  the barrier is quasi-static,  and the electron energy beneath the barrier always equals the binding energy $I_p$. However,  in the case of nonadiabatic processes,  the electron undergoes changes while tunneling,  resulting in the electron energy at the tunnel exit not being equal to $I_p$,  which is a key characteristic of nonadiabatic theory.

In recent years,  various experiments have been conducted to study the nonadiabatic effect in the ionization process {\cite{doi:10.1126/science.1163439,PhysRevLett.111.103003, PhysRevLett.105.133002, PhysRevLett.127.273201, PhysRevA.95.053425}}. Additionally, some analytical work has also been carried out to explore the nonadiabatic-related phenomenon {\cite{PhysRevA.98.013411, Boroumand_2022, Xiao:22}}. But in {\cite{PhysRevLett.111.103003}},  a contradictory conclusion was presented,  suggesting that for a wide range of intensities,  nonadiabatic theories proposed do not align with experimental data trends,  while adiabatic assumptions are supported. So further investigation into the ionization process is still necessary. Meanwhile,  recent studies have focused on HHG to analyze the yield in both adiabatic and nonadiabatic regions {\cite{2013Wavelength, 2012Strong}},  with some research concentrating on electron dynamics {\cite{2004Nonadiabatic, 2013High, 2023Revealing}}. However,  there is still a lack of comprehensive analytical work on the nonadiabatic effect in the HHG process, making it challenging to differentiate between the contributions of adiabatic and nonadiabatic effects on HHG spectra.

To calculate the HHG spectra, time-dependent Schr\"{o}dinger equation (TDSE) can be used. The numerical solution of the TDSE involving the Coulomb effect and can usually provide the accurate results. 
But even carried out with the single active electron (SAE) approximation, the TDSE is still computational demanding. To understand the complex mechanisms of HHG, the SFA method was proposed {\cite{Lewenstein94}} and has since been widely adopted {\cite{Weissenbilder2022, doi:10.1080/09500340412331301542, Xie:25}}. 
Under the assumptions of considering only the ground state in bound states, neglecting the depletion of the ground state, and ignoring the Coulomb effect, the SFA method can qualitatively describe the HHG and provide a clear physical picture of HHG. It is especially effective in the case of low-frequency driving lasers with a small Keldysh parameter {\cite{Keldysh65}}. 

Based on the SFA method, the description of HHG spectra mainly relies on numerical techniques such as multiple numerical integration {\cite{Lewenstein94}} or the coherent superposition of quantum trajectories, obtained by numerically solving the saddle-point equations {\cite{Lewenstein94,Le_2016,PhysRevLett.99.253903,ZAIR2013184}}. Consequently, investigations into the physical mechanisms of HHG are predominantly based on numerical results. The saddle-point solutions provided by SFA offer a clearer physical picture of HHG and can be easily extended to account for macroscopic propagation, benefiting from the method’s computational efficiency {\cite{Weissenbilder2022,PhysRevA.61.063801,Li2025}}. Therefore, the SFA method is an important approach for studying HHG. However, due to the three underlying assumptions of SFA, its results are often qualitative, although they are still able to capture the main physics of HHG observed in experiments or TDSE calculations {\cite{PhysRevA.91.023428,PhysRevA.78.023814,Li2025}}.


To improve the accuracy of SFA, the quantitative rescattering (QRS) theory \cite{PhysRevLett.100.013903,PhysRevA.80.013401} was proposed. In QRS, the HHG dipole is expressed as a function of the harmonic photon energy and is factorized into the product of the returning electron wave packet and the exact photo-recombination transition dipole. This approach results in a good match with some experimental results. However, this method still requires numerical integration or numerical solving of saddle point equations, which are not fully analytical.
Furthermore, theoretically, one of the most important factors ignored in SFA is the Coulomb effect, hence semiclassical trajectory-based strong-field methods, such as trajectory-based Coulomb-SFA (TCSFA) {\cite{PhysRevLett.105.253002,Yan2013}}, Coulomb-Corrected Strong-Field Approximation (CCSFA) {\cite{PhysRevA.87.023418}}, quantum-trajectory Monte Carlo (QTMC) {\cite{PhysRevLett.112.113002}}, Coulomb Quantum-Orbit Strong-Field Approximation (CQSFA) {\cite{PhysRevA.92.043407}}, semiclassical two-step (SCTS) {\cite{PhysRevA.94.013415}} models have been developed to quantitatively describe the ionization. In these methods, the  electron trajectory of SFA {\cite{PhysRevA.51.1495}} for ionization is first obtained numerically, which gives the initial conditions of the tunneling electron at the tunnel exit, including initial velocity and position. Then  the initial conditions are used in numerical solution of Newton equation including both Coulomb force and electric-field force. Based on the methods, recently, a Coulomb-modified SFA (MSFA) {\cite{Xie:20,Xie_2022}} have also been developed to quantitatively describe the HHG. However, these methods still require numerically solving saddle-point equations or other numerical procedures to determine the initial conditions. On the other hand, in HHG-based attosecond measurement, one needs to deduce the ultrafast dynamical information of the system from the experimental observables such as the HHG spectra. Therefore, a concise and precise mapping between the required information and the observable is highly desired. Such mappings can be obtained from a fully analytical and accurate Coulomb-included HHG model. While constructing such a HHG model, a SFA-based and fully analytical description of the HHG electron trajectories, instead of the numerical solution of saddle-point equation, is first needed.

To provide deeper insights into the nonadiabatic effect in HHG and explore an analytical expression for HHG electron trajectories, in the paper, we develop a fully analytical approach to describe HHG based on SFA and perturbation expansion theory. This provides an alternative method for the numerical solution of saddle-point equations in SFA (NSFA). Our method allows for a clear and analytical separation of adiabatic and nonadiabatic effects, thereby offering a more transparent physical picture of HHG and demonstrating that nonadiabatic effects can enhance the HHG yield. Furthermore, this approach provides analytical initial conditions for the electron at the tunnel exit, which influence subsequent electron dynamics and the resulting HHG yield. In fact, our fully analytical method is developed based on the SFA from a different perspective and does not conflict with other SFA-based methods mentioned above. Therefore, it is expected that our method can be combined with semiclassical trajectory-based Coulomb models or QRS model to achieve a full description of HHG in the future.


The structure of this paper is as follows. In Sec.~\ref{theo}, the fully analytical method based on SFA and the saddle point method is introduced. In Sec.~\ref{nonadiabatic}, the accuracy of the analytical method is validated by comparing its results with those obtained from NSFA method and the nonadiabatic effect on HHG is discussed. In Sec.~\ref{velocity and position}, the analytical nonadiabatic initial conditions for the electron at the tunnel exit are introduced. These conditions are essential for the semiclassical trajectory-based Coulomb-included strong-field models. In Sec.~\ref{comparisons}, we further discuss the significance of the work and compare it to other methods. Finally, in Sec.~\ref{concl}, we present our conclusions.

\section{Analytical description\label{theo}}

According to SFA,  when an atom is driven by the laser field $\textbf{E}(t)$,  the component of the time-dependent dipole moment along an arbitrary direction $\textbf{n}$ can be expressed as {\cite{Lewenstein94}}.

\begin{equation}
	\begin{aligned}
{D}_\textbf{n}(t)= & i\int_{0}^{t}dt' \int d^3 \textbf{p} \  \textbf{n} \cdot \textbf{d}^*_r \left(\textbf{p}-\textbf{A}(t)\right)  \\ \times &\textbf{E}(t') \cdot  \textbf{d}^*_i \left(\textbf{p}-\textbf{A}(t')\right) exp \left[-iS\left(\textbf{p}, t, t'\right)\right]+c.c. \\
    \end{aligned}
    \label{eq:time-dependent dipole moment1}
\end{equation}
Here,  $\tau \equiv t-t'$  representing the excursion time of the ionized electron,  with $t'$ and $t$ being the ionization and return time respectively. The vector potential of the field is denoted as $\textbf{A}(t)$,  satisfying the relation $\textbf{A}(t)\equiv -\int_{-\infty}^{t}\textbf{E}(t'')dt''$. The canonical momentum is denoted as $\textbf{p}$. In Equation (\ref{eq:time-dependent dipole moment1}),  $\textbf{d}_i \left(\textbf{p}-\textbf{A}(t')\right)$ represents the dipole transition matrix element between the ground state and the continuum state $|{\textbf{p}-\textbf{A}(t')}\rangle$. Similarly,  $\textbf{d}^*_r \left(\textbf{p}-\textbf{A}(t)\right)$ represents that from the continuum state $|{\textbf{p}-\textbf{A}(t)}\rangle$ to the ground state. $S$ is the quasiclassical action {\cite{Lewenstein94}},  describing the free electron motion driven by the laser field and has the expression:
\begin{equation}
	S=\int_{t'}^{t}{\left(\frac{\left[ \textbf{p}-\textbf{A}\left( t'' \right) \right] ^2}{2}+I_p\right) dt'' }. 
	\label{eq:quasiclassical action.}
\end{equation}

The primary contribution to the integral in Eq. (\ref{eq:time-dependent dipole moment1}) arises from the saddle points of the action. With respect to $\textbf{p}$, one can derive a saddle point equation:
\begin{equation}
	\begin{aligned}
		\nabla_{\textbf{p}} S  =   \int_{t'}^{t}{\left[ \textbf{p}-\textbf{A}\left( t'' \right) \right] dt'' }=0.
	\label{eq:saddle point equationa1}
	\end{aligned}
\end{equation}
Then one can obtain the saddle point canonical momentum 
\begin{equation}
	\begin{aligned}
	\textbf{p}_s = \frac{1}{t-t'}\int_{t'}^{t} \textbf{A}(t'')dt''.
	\label{eq:saddle point equationa2}
	\end{aligned}
\end{equation}
Assuming that the laser is linearly polarized along the $x$-axis, the saddle point canonical momentum is along the $x$-axis as well. With the saddle point method, Eq. (\ref{eq:time-dependent dipole moment1}) can be simplified to a one-dimensional form:
\begin{equation}
	\begin{aligned}
{D}_{x}(t)= & -i\int_{-\infty}^{t}dt' \big[\xi(\tau){d}_{r}^{*}\left({p}_s-{A}\left(t\right)\right)\\ 
& \times{E}(t')\cdot {d}_{i}\left({p}_s-{A}\left(t'\right)\right) \\
&\times e^{-iS({p}_s, t, t')}\big]+c.c. .\\
    \end{aligned}
    \label{eq:time-dependent dipole moment}
\end{equation}
The factor $\xi(\tau)=[\pi/(\epsilon'+i\tau/2)]^{\frac{3}{2}}$ arises from the Gaussian integration around the saddle points,  where $\epsilon'$ is an infinitesimal quantity used to prevent numerical singularities. $\xi(\tau)$ is associated with the spread of the electronic wave packet. Then the Fourier transform of ${D}_{x}(t)$ has the form {\cite{Le_2016}}
\begin{equation}
\begin{aligned}
	{D}_{x}(\Omega)=& -i\int dt \int_{-\infty}^{t}dt'\big[\xi(\tau) {d}_{r}^{*}({p}_s-{A}(t))\\ &\times {E}(t').
	{d}_{i}({p}_s-{A}(t'))\\
	&\times e^{-i \Theta(p_s, t, t')} \big].
\end{aligned}
\label{eq:frequency-dependent dipole moment1}
\end{equation}
In the above equation, the exponential term $\Theta(p_s, t, t')$ has the following form:
\begin{equation}
	\begin{aligned}
		\Theta(p_s, t, t')=S(p_s, t, t')-\Omega {t}/{\omega}.	
	\end{aligned}
	\label{eq:theta}
	\end{equation}
Here,  $\Omega$ represents the photon energy of the HHG,  which can be expressed as the sum of the kinetic energy of returned electron and the binding energy of the target atom. Similarly,  using Eq. (\ref{eq:saddle point equationa1}),  one can derive the other two saddle point equations from $\Theta$ with respect to $t'$ and $t$:
\begin{equation}
	\frac{\left[ p_s-A\left( t' \right) \right]^2}{2}+I_p=0, 
	\label{eq:saddle point equationb}
\end{equation}
	
\begin{equation}
	\frac{\left[ p_s-A\left( t \right) \right] ^2}{2}-\left(\Omega-I_p\right)=0.
	\label{eq:saddle point equationc}
\end{equation}
By solving the two equations above,  one can obtain a pair of saddle points $(t_r, t_i)$ for a given photon energy  $\Omega$. The time pair is also called the electron trajectory or the quantum trajectory of HHG, including long trajectory, short trajectory of the first return of the rescattering electron to the nucleus and multiple returns {\cite{Lewenstein94}}.
In the above expression, $t_r$ represents the saddle point return time and $t_i$ represents the saddle point ionization time. According to the electron trajectory theory, the induced dipole of a certain trajectory can be denoted as $D_{xj}(\Omega)$. The laser-induced time-dependent dipole moment along the laser polarization {\cite{Le_2016}} for the electron trajectory can be expressed as
\begin{equation}
	\begin{aligned}
		D_{xj}(\Omega)=&\xi(\tau)\frac{2\pi}{\sqrt{det(S'')}}{d}_{r}^{*}({p}_s-{A}(t_r)) \\ &\times {E}(t_i){d}_{i}({p}_s-{A}(t_i))e^{-i\Theta(p_s, t_r, t_i)}.
	\end{aligned}
	\label{eq:frequency-dependent dipole moment2}
\end{equation}
In this context,  the subscript ``j'' represents different trajectories,  such as ``S1'' refers to the short trajectory in the first return,  ``L1'' refers to the long trajectory in the first return,  ``S2'' refers to the short trajectory in the second return and so forth. Then the intensity of the HHG spectra can be expressed as 
\begin{equation}
	\begin{aligned}
		P(\Omega)\propto\left|\sum_{j}D_{xj}(\Omega)\right|^2.
	\end{aligned}
	\label{eq:P}
\end{equation}

According to SFA,  the three saddle point equations determine three key quantities: $p_s$,  $t_{r}$ and $t_i$ which play a dominating role in the HHG process. Based on the theory outlined above,  a detailed form of the laser field can be introduced to obtain the fully analytical results.

The linearly polarized laser field can be assumed to be a monochromatic field,  where ${E}(t)=E_0\cos(t)$. Here,  the independent variable $t$ represents the phase and will be used in the following text. Similarly,  time-related variables such as electron return time,  ionization time,  and excursion time are presented in this form. By utilizing Eq. (\ref{eq:saddle point equationa2}) and the expression ${E}(t)=E_0\cos(t)$,  one can easily derive the following expression for the canonical momentum:
\begin{equation}
	\begin{aligned}
		p_s=A_0 \cdot \frac{\left[{\cos (t_r)-\cos (t_r-\tau )}\right]}{{\tau }}.
	\end{aligned}
	\label{eq:ps(t)}
\end{equation}
Here,  $A_0$ satisfies the relation $A_0=2 \sqrt{U_p}$. Similarly,  the adiabatic parameter $\gamma$ mentioned in the first section can also be written as a function of $U_p$,  such that $\gamma=\sqrt{I_p/2U_p}$. The relationship between $U_p$ and $\gamma$ suggests that $U_p$ is a suitable scaling quantity of energy when expanding quantities with respect to $\gamma$,  accordingly $A_0$ is a suitable scaling quantity of momentum.
By scaling with $U_p$ and utilizing Eq. (\ref{eq:ps(t)}),  Eq. (\ref{eq:saddle point equationb}) can be simplified to a concise and equivalent form that expresses the relationship between $t_r$ and $\tau$ (the detailed derivation process can be found in appendix A), namely 
\begin{equation}
	\sin\left( t_r-\frac {\tau}{2}\right) a -\cos\left( t_r-\frac {\tau}{2}\right) s =i\gamma.
	\label{eq:saddleb reduced form}
\end{equation}
Here $a$ and $s$ are defined as
\begin{equation}
	\begin{aligned}
		&a\equiv a(\tau)= \cos(\tau/2)- \frac{ 2\sin(\tau/2)}{\tau},  \\
		&s\equiv s(\tau)= \sin(\tau/2).
	\end{aligned}	
	\label{eq:as}
\end{equation}
Similarly,  scaling with $U_p$ and utilizing Eqs. (\ref{eq:ps(t)}) and (\ref{eq:saddleb reduced form}),  equation (\ref{eq:saddle point equationc}) can also be simplified, and the kinetic energy of the returned electron can be expressed as a function of $\tau$ (the detailed derivation process can be found in appendix B), namely 
\begin{equation}
	\begin{aligned}
	E_{re}=&\frac {\left(p_s-A\left( t_r\right)  \right) ^2}{2} \\
	=& 2U_p[a \frac{ i a \gamma +\left(s \sqrt{a^2+s^2+\gamma ^2}\right)}{a^2+s^2} \\&+ s \frac{ -i s \gamma+ \left(a \sqrt{a^2+s^2+\gamma ^2}\right)}{a^2+s^2}]^2 .
	\end{aligned}
	\label{eq:kinetic energy}
\end{equation}
Here,  the kinetic energy of the returned electron satisfies the relationship $E_{re}=\Omega-I_p$. According to Eq. (\ref{eq:as}) and Eq. (\ref{eq:kinetic energy}),  $E_{re}$ is a function of the excursion time $\tau$ only. Therefore, the time $\tau$ can be expanded into a third-order expansion:
\begin{equation}
	\begin{aligned}
		\tau^{(3)} = \tau_0+\tau_1\gamma+\tau_2\gamma^2+\tau_3\gamma^3, 
	\end{aligned}
	\label{eq:tau3orig}
\end{equation}
with $\tau=\tau^{(3)}+O(\gamma^4)$, and $\tau_0$ to $\tau_3$ are the coefficients of the zeroth to third-order perturbation expansion. For simplicity, the expansions mentioned here will have the same forms where the subscript numbers denote the expansion terms' order. By replacing $\tau$ with $\tau^{(3)}$ and considering the third-order expansion with respect to $\gamma$,  the kinetic energy of electron at the return time can be expanded to a third-order form:
\begin{equation}
	\begin{aligned}
	E_{re}^{(3)}= U_p \cdot \left(E_{re_0} +\gamma E_{re1}  +\gamma^2 E_{re2}  +\gamma^3 E_{re3} \right).
	\end{aligned}
	\label{eq:Ere}
\end{equation}
In experiments,  as an observable quantity,  the value of $E_{re}$ must be  real,  meaning that $\mathrm{Im}[E_{re}]=0$. By applying this condition to Eq. (\ref{eq:Ere}),  the coefficient of the second-order term is derived to be $\tau_2=0$,  and $\tau_1$ and $\tau_3$ in Eq. (\ref{eq:tau3orig}) are shown to be imaginary. To make the results clearer,  we will redefine $\tau_1$ and $\tau_3$ as the real coefficients of the imaginary first and third-order terms,  then the third-order expansion of the excursion time $\tau$ can be rewritten as
\begin{equation}
	\begin{aligned}
		\tau^{(3)} = \tau_0+i\tau_1\gamma+i\tau_3\gamma^3, 
	\end{aligned}
	\label{eq:tau3}
\end{equation}
where the coefficients of higher-order expansion terms, which are functions of $\tau_0$, take the following forms:
\begin{equation}
	\begin{aligned}
	\tau_1=& -\frac {\tau_0 \sqrt{a_0^2+s_0^2}} {2 a_0 s_0 + \tau_0 \left(a_0^2+s_0^2\right)}, \\
	\tau_3=&\frac{\tau_0 \sqrt{a_0^2+s_0^2}}{6 \left(a_0^2-s_0^2\right) \left(a_0^2 \tau_0+2 a_0 s_0+s_0^2 \tau_0\right)^4} \\ 
		&\times \big[a_0^6 \tau_0^3+2 a_0^5 s_0 \tau_0^2+a_0^4 s_0^2 \tau_0 \left(\tau_0^2-6\right)\\
		&-12 a_0^3 s_0^3-a_0^2 s_0^4 \tau_0 \left(\tau_0^2-20\right)\\
		&-2 a_0 s_0^5 \left(\tau_0^2-6\right)-s_0^6 \tau_0 \left(\tau_0^2-2\right)\big]. 
	\end{aligned}
	\label{eq:tau1}
\end{equation}
Here,  $a_0\equiv a(\tau_0)$ and $s_0\equiv  s(\tau_0)$. By inserting this expression of $\tau_3$ into Eq. (\ref{eq:kinetic energy}) again,  and expanding it to third order with respect to $\gamma$,  one can obtain the expression of $E_{re}^{(3)}$:
\begin{equation}
\begin{aligned}
	E_{re}^{(3)}= U_p \cdot (E_{re0}+E_{re2}{{\gamma}^2}).
	\end{aligned}
	\label{eq: third order kinetic energy}
\end{equation}
The coefficients of zero and second order expansions are {\cite{PhysRevA.106.053105}} 
\begin{equation}
	\begin{aligned}
	&E_{re0}= \frac{8a_0^2s_0^2}{a_0^2+s_0^2}, \\
	&E_{re2}=- \frac{ 16 a_0^2 s_0^3 \left( s_0+a_0\tau_0\right)  }{\left(a_0^2+s_0^2 \right)\left( 2a_0s_0+a_0^2\tau_0+s_0^2\tau_0\right) ^2 },
	\end{aligned}
	\label{eq:coefficients of E_re}
\end{equation}
while the first-order and third-order expansion terms are zero,  i.e.,  $E_{re1}=0$ and $E_{re3}=0$.

According to Eq. (\ref{eq:saddleb reduced form}), the expression of return time $t_r$ can also be written as a function of $\tau$  (the detailed derivation is shown in appendix B). Then,  $t_r$ can be expanded to third order directly with respect to $\gamma$. That is
\begin{equation}
	\begin{aligned}
	t_{r}^{(3)}=t_{r0}+ t_{r2}\gamma^2+i t_{r3}\gamma^3.	
	\end{aligned}
\label{eq:tr}
\end{equation}
where the coefficients of the expansions are functions of $\tau_0$ and have the forms of
\begin{equation}
	\begin{aligned}
		t_{r0} =& \tau_0/2 + \arctan(a_{0}, s_{0}), \\
		t_{r2} =& \frac{\tau_0 \cos \tau_0 - \sin \tau_0}{a_0^2+s_0^2}, \\
		t_{r3} =& \frac{2\sin ( \tau_0 / 2) ^2}{3(\tau_0-\sin\tau_0)^3}  \\
		&\times\frac{(2+\tau_0^2-2\cos\tau_0-2\tau_0 \sin\tau_0)^{3/2} }{[2+(\tau_0^2-2)\cos \tau_0-2\tau_0\sin\tau_0]}.
	\end{aligned}
\label{eq:tr coefficient}
\end{equation}

Based on Equations (\ref{eq:ps(t)}),  (\ref{eq:tau3}),  and (\ref{eq:tr}),  the canonical momentum $p_s$ can also be expanded to the third order:
\begin{equation}
	\begin{aligned}
		p_{s}^{(3)}=A_0 \cdot (p_{s0}+ p_{s2}\gamma^2+i p_{s3}\gamma^3), 	
	\end{aligned}
\label{eq:p_{st3}}
\end{equation}
where the coefficients of the expansions have the forms of
\begin{equation}
	\begin{aligned}
		p_{s0} =& -\sin(t_{r0} - \tau_0), \\
		p_{s2} =&-\frac{\tau_1^2 \cos (t_{r0}-\tau_0)}{2 \tau_0} \\
		& -\frac{t_{r2} \left[\sin (t_{r0})-\sin (t_{r0}-\tau_0)\right]}{\tau_0}, \\
		p_{s3} =& -\frac{1}{6(\tau_0)^2}[(6 \tau_0 \tau_1 t_{r2}-3 \tau_1^3) \cos(t_{r0}-\tau_0) \\
		&+6(\tau_0 t_{r3}-\tau_1 t_{r2}) \sin(t_{r0})\\
		&+(-6\tau_0 t_{r3}+3\tau_1 t_{r2}) \sin(t_{r0}-\tau_0)].
	\end{aligned}
\label{eq:p_s}
\end{equation}
Here the first order expansion term of $p_s$ is zero,  i.e.,  $p_{s1}=0$.
Based on the expansions mentioned earlier,  the saddle point excursion time $\tau$ and return time $t_{r}$ have been expanded to their third-order forms: $\tau^{(3)}$ and $t_{r}^{(3)}$. By using the relationship $t_i^{(3)}=t_r^{(3)}-\tau^{(3)}$,  the ionization time can be determined. The third-order ionization time can be expressed as
\begin{equation}
	\begin{aligned}
	t_{i}^{(3)}=t_{i0}+i t_{i1}\gamma+ t_{i2}\gamma^2+i t_{i3}\gamma^3.	
	\end{aligned}
\label{eq:ti}
\end{equation}

In our previous work {\cite{PhysRevA.106.053105}},  the second-order analytical electron kinetic energy $E_{re}$ was obtained, and the influence of the nonadiabatic effect on the cutoff energy of HHG was studied. In this section,  with respect to $\gamma$,  the third-order expansion of the three saddle point quantities $t_{r}^{(3)}$,  $t_{i}^{(3)}$ and $p_s^{(3)}$ has been performed as functions of $\tau_0$. In the following sections, the HHG spectra will be further examined based on the derivations of $t_{r}^{(3)}$,  $t_{i}^{(3)}$ and $p_s^{(3)}$. The applicability of the analytical method will then be discussed in detail, and the nonadiabatic effect on HHG spectra will be demonstrated.

\section{Nonadiabatic effect in HHG\label{nonadiabatic}}
According to the SFA,  the exponential factor $e^{-i\Theta}$ dominates the dipole moment Eq. (\ref{eq:frequency-dependent dipole moment2}). To demonstrate the applicability of this analytical method,  we will first discuss the exponential factor $\Theta$. This factor is critically important for the HHG yield as it is directly related to the ionization ratio \cite{Auguste_2012}. The variable $\Theta$ can be scaled by $U_p/\omega$,  allowing us to define a new variable $\theta$ that satisfies the relation:
\begin{equation}
	\begin{aligned}
		\Theta(p_s, t_r, t_i)=\frac{U_p}{\omega}\theta.	
	\label{eq:Theta}
	\end{aligned}
\end{equation}
Then one can expand $\theta$ to either third-order analytical expansion (TAE) or fifth-order analytical expansion (FAE) as a function of $\tau_0$ and they are

\begin{equation}
	\begin{aligned}
	&\theta^{(3)}=\theta_0(\tau_0)+\theta_2(\tau_0) \gamma ^2+i \theta_3(\tau_0) \gamma ^3,\\
	&\theta^{(5)}=\theta^{(3)}+\theta_4(\tau_0)\gamma ^4+i\theta_5(\tau_0)\gamma ^5.
	\label{eq:theta3}
	\end{aligned}
	\end{equation}
Based on our derivation,  the even-order expansion terms are real-valued,  but the odd-order terms are imaginary-valued.
Then the HHG contribution of a certain quantum trajectory can be expressed as 
\begin{equation}
	\begin{aligned}
		\vert{D}_{xj}(\Omega)\vert^2 &\propto e^{2 \mathrm{Im}\{\Theta\}  }\\ 
		&\propto \mathrm{exp}( \theta_3\frac{ I_p}{ \omega} \gamma)\mathrm{exp}( \theta_5\frac{ I_p}{ \omega}\gamma^3)\\
		&\propto \mathrm{exp}( \sqrt{2}\theta_3\frac{ I_p^{1.5}}{ I_0^{0.5}} )\mathrm{exp}( 2\sqrt{2}\theta_5\frac{4 \pi I_p^{2.5}}{I_0^{1.5} \lambda^2 }) \\
		&\propto \mathrm{exp}( 2\Theta_3)\mathrm{exp}( 2\Theta_5).
	\label{eq:ADK}
	\end{aligned}
\end{equation}
The imaginary parts of $\theta$ contribute directly to HHG amplitudes. Here the third-order scaling law and the fifth-order scaling law are derived. Here $E_0$ and laser intensity $I_0$ satisfy the relationship of $E_0^2\propto I_0 $, hence third-order scaling law is only determined by the binding energy and laser intensity, while the fifth-order scaling law is associated with the wavelength in addition. Theoretically,  one needs to at least the TAE to obtain the amplitude of spectra. However, further to improve accuracy, obtaining fourth- and fifth-order corrections is necessary. Because of the periodicity of the laser field, for simplicity, in the following discussions, unless mentioned otherwise, we only consider the results related to trajectories born at a single half laser cycle.
\begin{figure}
	\begin{center}
		\includegraphics[width=1\columnwidth]{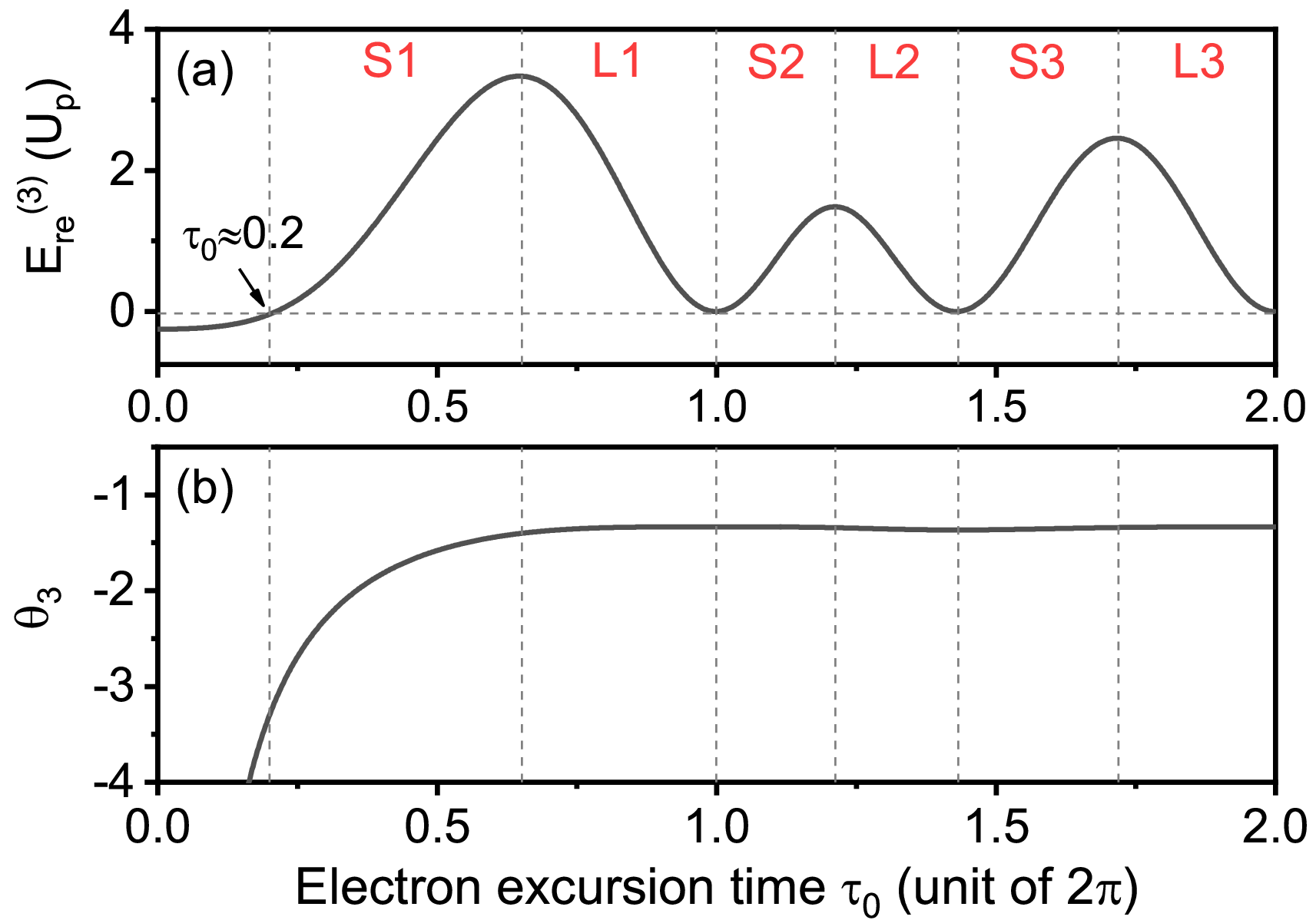}
	\end{center}
	\caption{{(a) The returned electron kinetic energy $E_{re}^{(3)}$ for $\gamma=0.5$ and (b) parameter-independent exponential factor coefficients $\theta_3$ as functions of the electron excursion time $\tau_0$.}} 
	\label{fig:theta}
\end{figure}

In Fig.~\ref{fig:theta} (a),  the third-order expansion of the returned electron kinetic energy,  $E_{re}^{(3)}$,  is plotted as a function of $\tau_0$. The figure is divided into different parts by vertical dashed lines to distinguish the different electron trajectories, and the short trajectories of the first return to the third return are denoted as S1, S2, S3, while the long trajectories of the first return to the third return are denoted as L1, L2, L3.
The horizontal dashed line represents the zero return energy. The intersection of the zero energy line and $E_{re}^{(3)}$ occurs at $\tau_0 \approx  0.2$,  indicated by the arrow. It is observed that $E_{re}^{(3)}$ is positive for almost every trajectory except for the S1 trajectory. In the case of the S1 trajectory,  $E_{re}^{(3)}$ is negative for $\tau_0< 0.2$,  indicating that the electron is extremely difficult to return to the core according to electron trajectory theory. According to the SFA,  $E_{re}^{(3)}$ being smaller than zero corresponds to below-threshold harmonics, and the accuracy of the approximation decreases. Therefore,  our analytical method is not suitable for describing the mechanism in this region,  and it will not be further discussed in this paper.

In Fig.~\ref{fig:theta} (b),  $\theta_3$,  represented by the black solid line,  is always negative for all trajectories and changes with $\tau_0$. For the S1 trajectory,  $\theta_3$ increases rapidly with increasing $\tau_0$,  potentially leading to a significant change in HHG yields. However,  for multiple returns and the L1 trajectory,  $\theta_3$ only oscillates slightly,  resulting in stable HHG yields.

In Eq. (\ref{eq:ADK}), the negative value of $\theta_3$ lays the basis for HHG yields $\mathrm{exp}( \sqrt{2}\theta_3\frac{ I_p^{1.5}}{E_0} )$ and the specific expression of $\theta_3$ and its physical meaning will be discussed later. Our derivation shows that $\theta_5$ is always positive,  so the fifth-order exponential term is always greater than 1, then $\mathrm{exp}( 2\sqrt{2}\theta_5\frac{I_p^{2.5}\omega^2}{E_0^3 })$ which is determined by the parameters of $I_p$, $E_0$ and laser frequency $\omega$ can be considered as a reinforced correction to the HHG yields and provide a more accurate result. 

\begin{figure}
	\begin{center}
		\includegraphics[width=1\columnwidth]{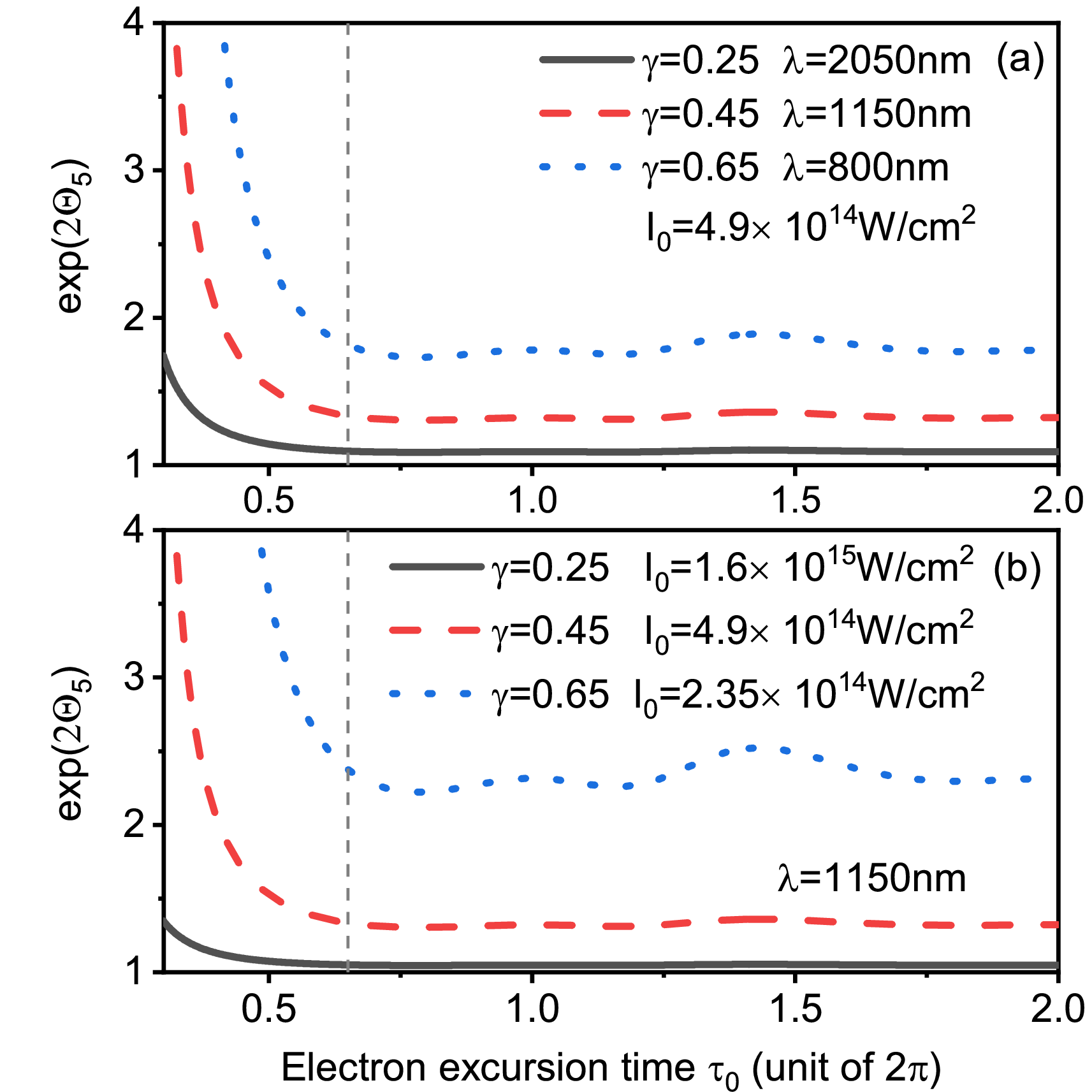}
	\end{center}
	\caption{ The higher-order factor $\mathrm{exp} (2\Theta_5)$ of HHG in Eq. (\ref{eq:ADK}) for different parameters. (a) $\mathrm{exp}(2\Theta_5)$ as a function of $\tau_0$ for various driving laser wavelength $\lambda$,  with $I_0=\mathrm{4.9 \times 10^{14}~W/cm^2}$. (b) $\mathrm{exp}(2\Theta_5)$ as a function of $\tau_0$ for various laser intensity $I_0$, with $\lambda  \mathrm{=1150~nm}$. The target atom is helium and the black solid lines, red dashed lines, blue short-dashed lines represent the results for $\gamma=0.25$, $0.45$, $0.65$ respectively.}
	\label{fig:theta5}
\end{figure}

Figure. \ref{fig:theta5} (a) illustrates the fifth-order correction as a function of $\tau_0$ and laser wavelength $\lambda_0$. It is evident that when the wavelength $\lambda$ decreases, $\mathrm{exp} (2\Theta_5)$ increases,  indicating that the fifth-order correction to HHG yield becomes more significant. The fifth-order correction factor $\mathrm{exp} (2\Theta_5)$ related to $\gamma$ is totally the contribution of nonadiabatic effect in HHG. For a laser intensity of $I_0=\mathrm{4.9\times10^{14}~W/cm^2}$,  when $\gamma=0.65$, the nonadiabatic effect can boost the HHG yield by over 150\%. 

In Fig. \ref{fig:theta5} (b),  the fifth-order correction  is plotted against $\tau_0$ and laser intensity $I_0$. Similar with Fig. \ref{fig:theta5}(a), the decrease in laser intensity results in a larger fifth-order correction. For a laser intensity of $\lambda\mathrm{=1150~nm}$, when $\gamma=0.65$, the fifth-order correction increases the HHG yield more than 200\%. 

The boundary between S1 and L1 trajectories is indicated by a vertical dashed line in Fig.~\ref{fig:theta5}. Generally, the fifth-order correction remains stable for L1 and other multiple return trajectories, but for the S1 trajectory (region where $\tau_0< 0.65$), the fifth-order correction increases significantly with decreasing $\tau_0$. And the nonadiabatic correction is always larger than 1, indicating that the nonadiabatic effect always enhances the HHG yield. Compared to the results of $\gamma=0.45$, the enhancement of $\gamma=0.65$ for shorter laser wavelength is more pronounced than that for smaller laser intensity as the comparison between Figs.~\ref{fig:theta5} (a) and (b) shows. The increase of the HHG yield with the decrease of the laser wavelength induced by the nonadiabatic effect relating to the fifth order expansion provides deeper insights into the wavelength scaling of HHG yields. It is well known that the HHG yield decreases with increasing the laser wavelength and shows a scaling of $\lambda^{-(5-6)}$ 
\cite{PhysRevA.65.011804,PhysRevLett.98.013901,PhysRevLett.103.073902,Li_2016}. 
The potential mechanism is mainly attributed to the wavepacket spreading. The wavelength-dependent scaling law of HHG yields is important for obtaining bright attosecond pulses from HHG in experiments. Our results in Fig. \ref{fig:theta5} obtained from the fully analytical HHG model indeed show that the nonadiabatic effect also plays a role in the scaling of HHG yields with respect to the laser wavelength.

In order to qualitatively determine the applicability of TAE and FAE in simulating the HHG spectra, we introduce the concept of error rate expressed as: 
\begin{equation}
	\delta =\frac{\left\lvert P_{j}^{(TAE/FAE)}(\Omega_c)-P_{j}^{(NSFA)}(\Omega_c) \right\rvert }{P_{j}^{(NSFA)}(\Omega_c)}. 
	\label{eq:error rate}
\end{equation}
In the equation above, the error rate $\delta$ of HHG intensity is a relative error. It is defined as the ratio of the difference between NSFA and TAE/FAE results over NSFA results.
We mention that in the calculation of HHG intensity using Eq. (\ref{eq:P}), we have used the full expression of Eq. (\ref{eq:frequency-dependent dipole moment2}) for the dipole moment where the dipole transition matrix elements $d_r$ and $d_i$ are included. Specifically, we expand these terms $t_i, t_r, \tau , p_s$ and $\Theta$ in Eq. (\ref{eq:frequency-dependent dipole moment2}) according to the TAE or the FAE method. Then the value of $D_{xj}$ of Eq. (\ref{eq:frequency-dependent dipole moment2}) related to the third-order expansion or the fifth-order expansion can be analytically obtained. Similar treatments are also used for obtaining the results in Fig. \ref{fig:HHG}.
In addition, for the region where the energy is greater than the HHG cutoff energy, the calculation of NSFA for the S1 trajectory will diverge unphysically {\cite{Auguste_2012}}, so the photon energy $\Omega_c$ chosen here is slightly lower than the cutoff energy to avoid numerical divergence. 

\begin{figure}
	\begin{center}
		\includegraphics[width=1\columnwidth]{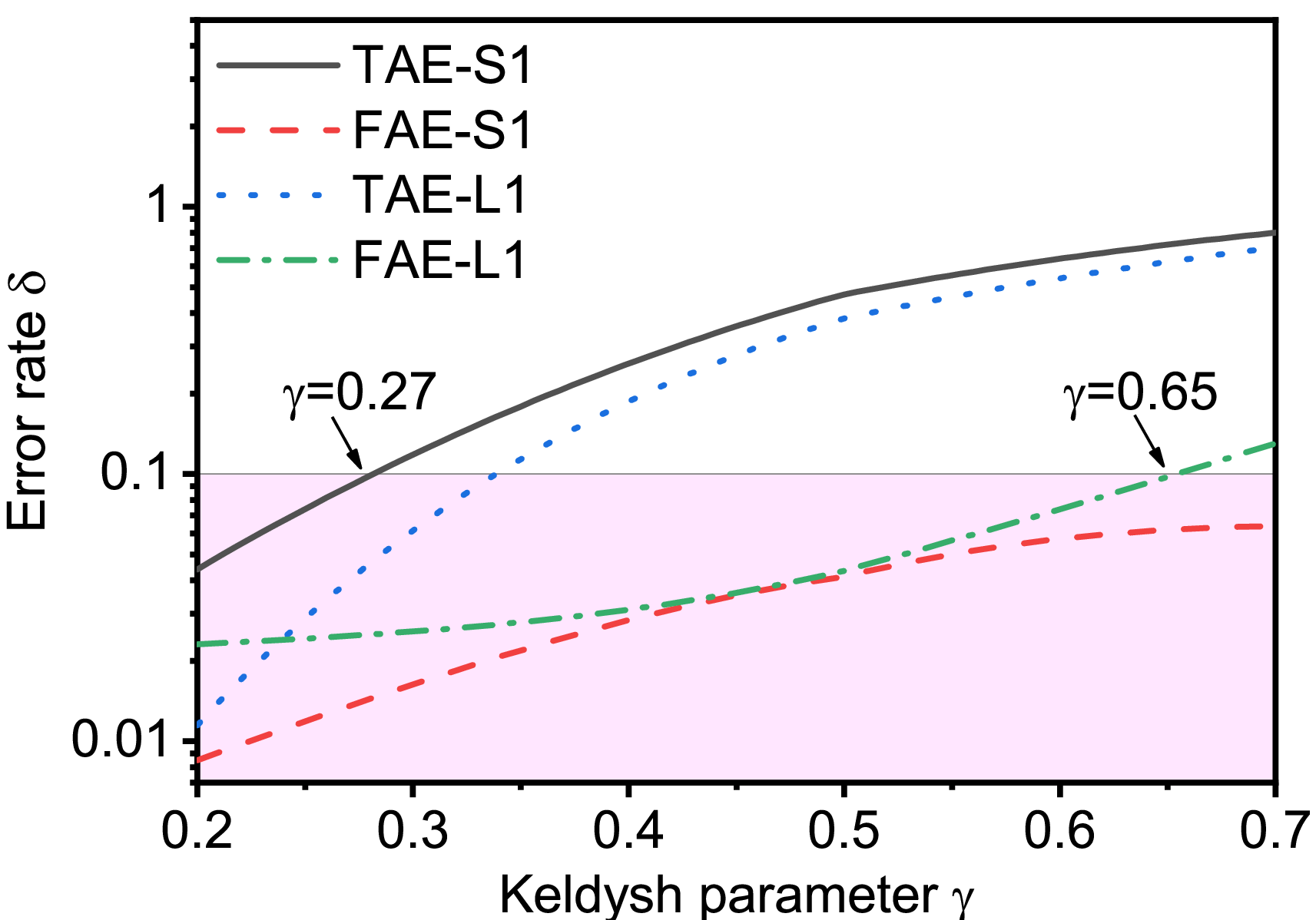}
	\end{center}
	\caption{Error rates of TAE and FAE methods for short and long trajectories of the first return as a function of the adiabatic parameter $\gamma$. The laser wavelength is $\lambda\mathrm{=1600~nm}$, and the intensity ranges from $\mathrm{1.05\times 10^{14}~W/cm^2}$ to $\mathrm{1.28\times 10^{15}~W/cm^2}$, corresponding to the range of $0.7\geqslant \gamma \geqslant 0.2$.}
	\label{fig:error rate}
\end{figure}
In Fig.~\ref{fig:error rate}, the error rates of the TAE and FAE methods are plotted against the adiabatic parameter $\gamma$. The error rates $\delta$ of the TAE method for the S1 and L1 trajectories are depicted by the black solid and blue short-dashed lines, respectively, while the error rates of the FAE method for the S1 and L1 trajectories are shown by the red dashed and green dash-dotted lines, respectively. It is set that when $\delta$ is less than 0.1, the analytical calculation of HHG spectra is acceptable, corresponding to the magenta area in Fig.~\ref{fig:error rate}. One can observe from Fig.~\ref{fig:error rate}, for the TAE method, it is applicable when $\gamma<0.27$, while for the FAE method, it is applicable when $\gamma<0.65$.

In the HHG process, the nonadiabatic effect will first impact the ionization yield, then influences the following electron propagation and return processes. Finally, the nonadiabatic effect will be retained in the HHG yield. Next, we give a detailed discussion to the above points.

Based on the TAE method, $\theta_3$ in Eq. (\ref{eq:theta3}) and the corresponding HHG yield related to the third-order expansion can be expressed as 
\begin{equation}
	\theta_{3}=-\frac{4}{3\cos t_{i0}},
	\label{eq:theta_3}
\end{equation}
\begin{equation}
	\begin{aligned}
		\vert{D}_{xj}(\Omega)\vert^2 &\propto \mathrm{exp}( \theta_3\frac{ I_p}{ \omega} \gamma)\\
		&\propto \mathrm{exp}( -\frac{2 (2 I_p)^{1.5}}{3 E_0 \cos t_{i0}}),
	\end{aligned}
\end{equation}
\label{eq:ADK31}
and the imaginary part of the exponential factor $\Theta$ is given by
\begin{equation}
	\mathrm{Im}[\Theta] \approx \Theta^{(3)}=\Theta_{3}= -\frac{ U_p (2I_p)^{1.5}}{3 \omega E_0 \cos t_{i0}} .
	\label{eq:ADK32}
\end{equation} 
Here, the third-order term $\theta_3$ is solely determined by $t_{i0}$, which is the zero-order coefficient of ionization time $t_i$ as in Eq. (\ref{eq:ti}). It suggests that in the TAE method, the HHG yield is determined by the ionization process only, which agrees with ADK theory in adiabatic approximation.

Based on the FAE method which includes higher-order terms, the HHG yield related to the fifth-order expansion can be expressed as
\begin{equation}
	\begin{aligned}
		 \vert{D}_{xj}(\Omega)\vert^2 & \propto
		 \mathrm{exp}( 2\Theta^{(3)}+2\sqrt{2}\theta_5\frac{I_p^{2.5}\omega^2}{E_0^3 }).
			\label{eq:ADK51}
	\end{aligned}
\end{equation}
and the imaginary part of the exponential factor $\Theta$ is given by
\begin{equation}
	\mathrm{Im}[\Theta] \approx \Theta^{(5)}= \Theta_{3} + \sqrt{2}\theta_5\frac{I_p^{2.5}\omega^2}{E_0^3 }.
	\label{eq:ADK52}
\end{equation}
Here an extra dependence on laser frequency $\omega^2$ is introduced, and this result can then be considered as a nonadiabatic one. The fifth-order coefficient $\theta_5$ in Eqs. (\ref{eq:ADK51}) and (\ref{eq:ADK52}) is cumbersome and relies on both the ionization time $t_i$ and the return time $t_r$ (see Appendix C). When considering the fifth-order expansion, the HHG yield is not only determined by the ionization process, but also the propagation and return processes, leading to a more accurate result.

\begin{figure}
	\begin{center}
		\includegraphics[width=1\columnwidth]{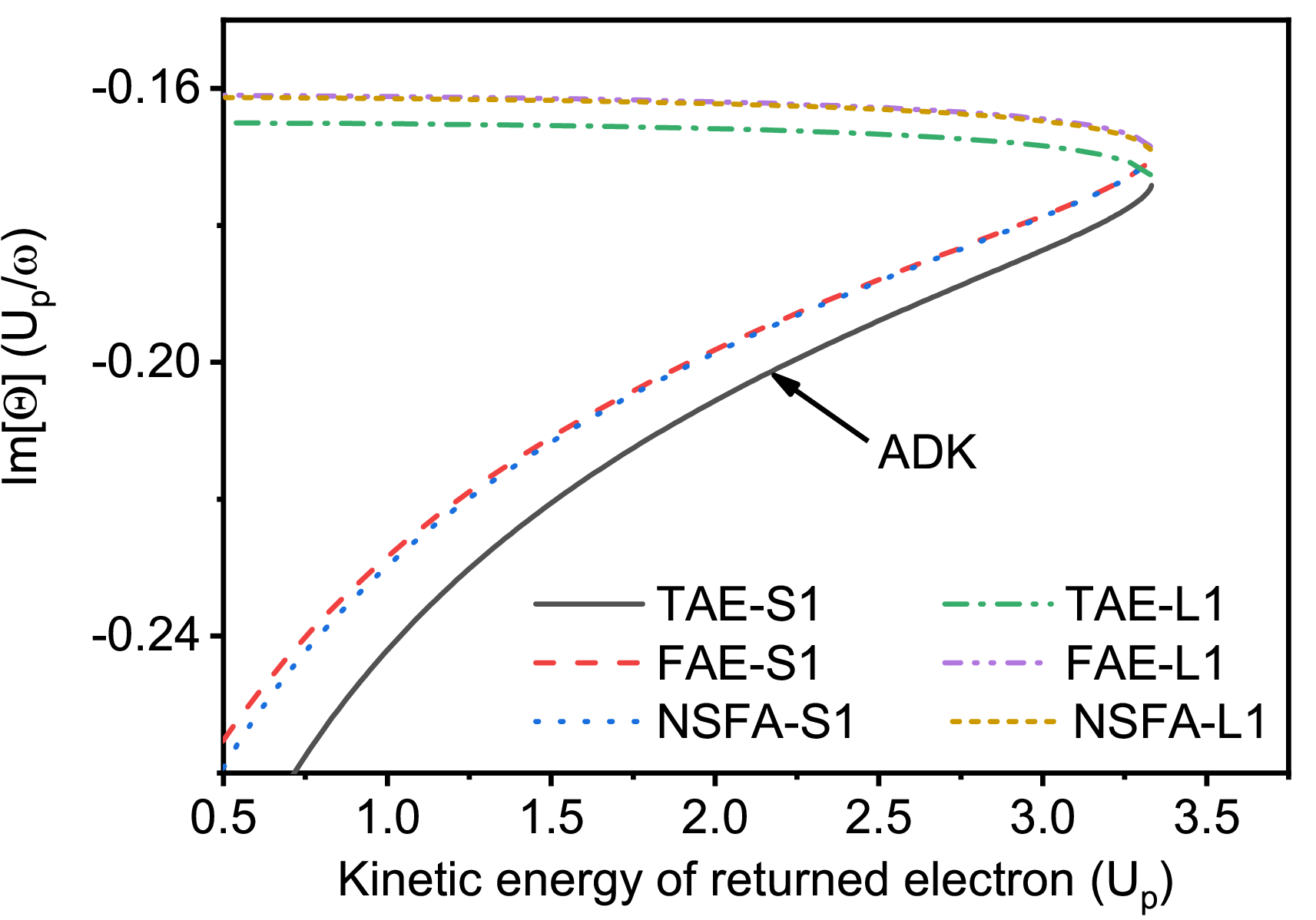}
	\end{center}
	\caption{ The exponential factor $\mathrm{Im}[ \Theta] $ as a function of electron return kinetic energy $E_{re}$ obtained using different methods when $\gamma=0.5$. The results for the S1 and L1 trajectories obtained by the NSFA method are shown in black solid and red long-dashed lines, while the results from of Eq. (\ref{eq:ADK32}) are represented by blue dotted and purple dash-dotted lines for the S1 and L1 trajectories, respectively. The results obtained by the Eq. (\ref{eq:ADK52}) are shown in green dash-dot-dotted and brown short-dashed lines for the S1 and L1 trajectories, respectively.} 
	\label{fig:ADKcompare}
\end{figure}
As shown in Eq. (\ref{eq:ADK}), the $\mathrm{Im}[\Theta]$ plays an important role in HHG yields. So in Fig.~\ref{fig:ADKcompare}, a comparison of imaginary part of the exponential factor $\Theta$ obtained with different methods is presented. 
The NSFA is used as a reference, indicated by the black solid and red long-dashed lines, while the results of TAE-S1 and TAE-L1 are represented by the blue dotted and purple dash-dotted lines, respectively. 
When comparing the NSFA and TAE results, a significant difference is observed at low electron return energy regions for S1 trajectory.
This is likely due to the later tunneling-out time of the short trajectory in half a laser cycle, corresponding to lower instantaneous electric fields at the exit time $t_{ex}=\mathrm{Re}[t_i]\approx t_{i0}$ {\cite{PhysRevLett.93.233002, PhysRevA.96.023406}}. 
Consequently, there is a larger effective adiabatic parameter $\gamma(t_{ex})=\omega \sqrt{2I_p} /E(t_{ex})$ {\cite{Auguste_2012}}, indicating a greater nonadiabatic effect. 
However, as the kinetic energy of returned electron increases, the difference between TAE and NSFA results becomes smaller.
On the other hand, in comparison with TAE-S1, the prediction of TAE-L1 is closer to the corresponding NSFA result. This is due to a smaller effective adiabatic parameter $\gamma(t_{ex})=\omega \sqrt {2I_p} /E(t_{ex})$, indicating a reduced nonadiabatic effect. 
The TAE curves can also be plotted using Eq. (107) in Ref. {\cite{Le_2016}}, where the exponential factor is equal to $-(2I_p)^{3/2}/(3E(t_{ex}))$, in an ADK-like form when assuming the laser field is static.

Compared to TAE results, both the FAE-S1 and FAE-L1 curves agree better with NSFA results in all energy regions. The inclusion of high-order terms in the calculations significantly improves the accuracy of HHG amplitude, especially in low-energy region with larger nonadiabatic effects. It is evident that FAE yields are always larger than TAE yields. This can be attributed to the nonadiabatic effect which increases the HHG amplitude.

As shown in Eq. (\ref{eq:P}), the HHG yield is a coherent sum of contributions from different quantum trajectories, and the interference between them must be included. The accuracy of the analytical method for describing interference effects in HHG is further confirmed in Fig.~{\ref{fig:HHG}}. 

\begin{figure}
	\begin{center}
		\includegraphics[width=1\columnwidth]{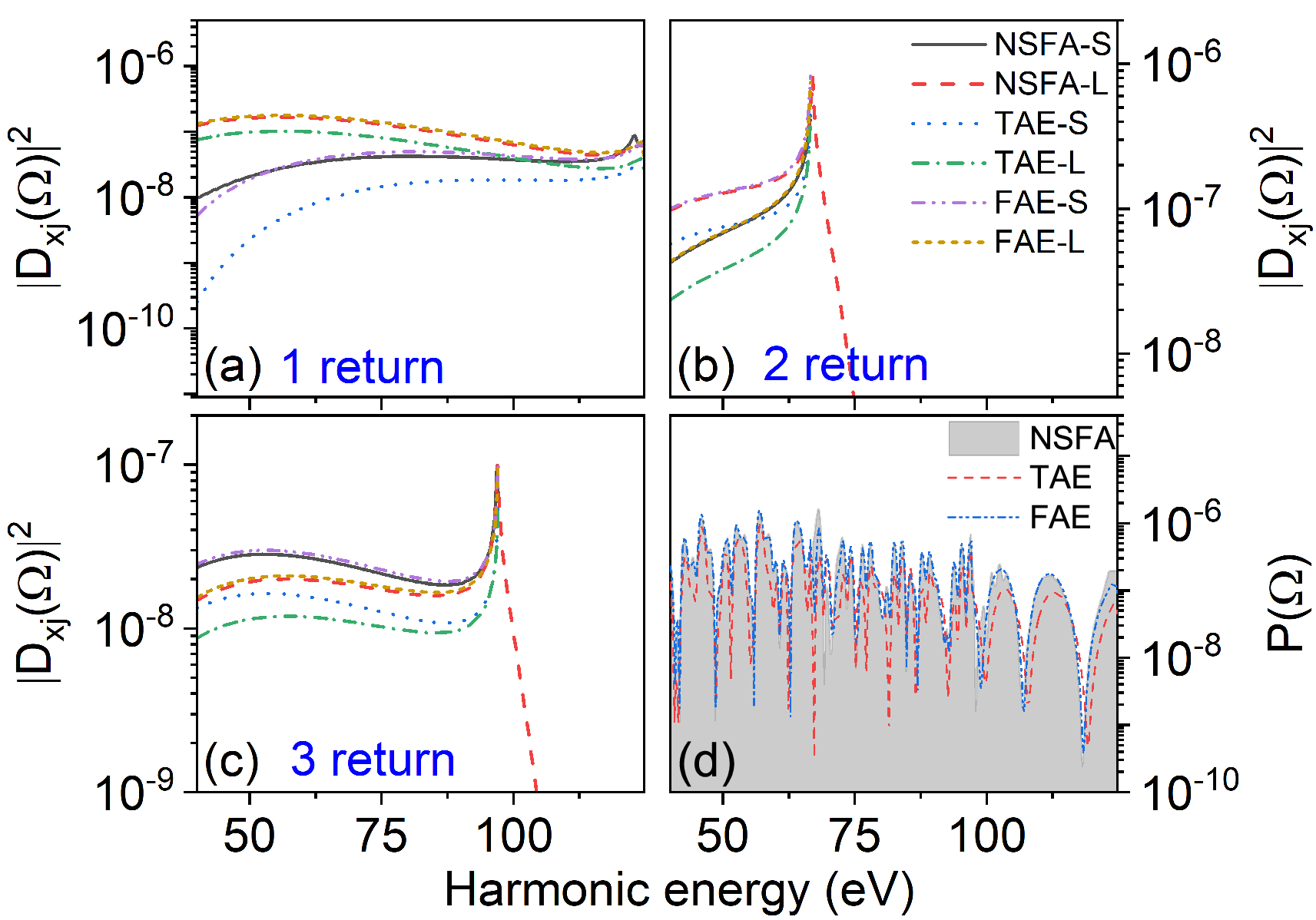}
	\end{center}
	\caption{
	Contributions of long (L) and short (S) trajectories of the first return (a), the second return (b) and the third return (c)  to the HHG obtained with Eq.~(\ref{eq:frequency-dependent dipole moment2}), and the HHG spectrum (d) involving  contributions of all trajectories of these three returns obtained with Eq.~(\ref{eq:P}).  
	The results are for the helium atom driven by a laser field with intensity of $\mathrm{I=5\times 10^{14}~W/cm^2}$ and wavelength of $\lambda$ = 790~nm, corresponding to the parameter $\gamma=0.65$.}
	\label{fig:HHG}
\end{figure}


In Figs.~{\ref{fig:HHG}} (a) to (c), we display the HHG contributions from the first, second, and third return trajectories. The black solid and red dashed lines represent the contributions from the short (S) and long (L) trajectories obtained using the nonadiabatic NSFA method. The blue dotted and green dash-dotted lines represent the short and long trajectories obtained using the adiabatic TAE method. The purple double-dotted dashed line and brown short-dashed lines represent the short and long trajectories obtained using the nonadiabatic FAE method, respectively. When comparing the NSFA and TAE methods, the HHG yields predicted by TAE are lower than those of NSFA for all three returns, particularly in the low-energy region of the S1 trajectory. However, the FAE method, which incorporates the fifth-order correction factor \(\mathrm{exp}(2\Theta_5)\), performs significantly better, resulting in HHG yields that closely match those of NSFA for all three returns. The discrepancy between the TAE and FAE results aligns with Fig.~\ref{fig:theta5}, suggesting a nonadiabatic enhancement in HHG. 



The HHG spectrum, which includes contributions from the three return trajectories simulated using the NSFA method (represented by gray shading), the TAE method (illustrated with red dashed lines), and the FAE method (shown with short-dashed lines), is presented in Fig.~{\ref{fig:HHG}}(d). One can observe that for the case of $\gamma=0.65$, when comparing the TAE results with the NSFA results, there are noticeable deviations. In contrast, even when considering the contributions of the first three return trajectories to HHG, the FAE method accurately reproduces the HHG spectrum obtained from the NSFA. It successfully captures both the harmonic yield and the interference structure, which is particularly evident in the high-energy region near the cutoff.
Considering the accuracy of FAE shown above, with the use of the accurate transition dipole matrix in Eq. (\ref{eq:frequency-dependent dipole moment2}), the fully-analytical FAE method can also be easily generalized to the QRS. 

\section{Nonadiabatic initial velocity and position at the tunnel exit\label{velocity and position}}
Different from adiabatic tunneling, the non-adiabatic tunneling ionization process involves various dynamics beneath the barrier. Ultimately, the electron emerges from the tunnel exit with a non-zero velocity and position \cite{PhysRevA.98.013411,PhysRevLett.93.233002,PhysRevA.96.023406}. The initial velocity and position associated with different harmonic orders indicate distinct dynamic processes occurring under the barrier. Furthermore, as mentioned in Section \ref{intro}, some trajectory-based numerical Coulomb-corrected SFA models proposed earlier \cite{PhysRevLett.105.253002,Yan2013,PhysRevA.87.023418,PhysRevLett.112.113002,PhysRevA.92.043407,PhysRevA.94.013415,Xie:20} have demonstrated a strong dependence on the initial conditions of the electron at the tunnel exit, which determine the electron's dynamics after emerging from the barrier. Our fully analytical method provides a way to derive these initial conditions without resorting to numerical manners, offering the possibility for developing fully-analytical  Coulomb-corrected quantitative SFA models.

Next, we discuss the analytical expressions of these initial conditions. Based on the SFA theory, the velocity of an electron at the tunnel exit is determined by its motion under the barrier. Consequently, nonadiabatic effects play a significant role. According to the analytical method, the velocity at the tunnel exit which is complex can be expressed as $v_{ex}=p_s-A(t_{ex})$ and can also be expanded as 
\begin{equation}
	\begin{aligned}
	v_{ex}\approx& p_{s}^{(5)}-A_{ex}^{(5)} \\
	\approx& p_{s0}+p_{s2}\gamma^2 +p_{s4}\gamma^4 -A_{ex0}-A_{ex2}\gamma^2 -A_{ex4}\gamma^4 \\
	&+i(p_{s3}\gamma^3 +p_{s5}\gamma^5)		
	\end{aligned}
\label{eq:v(t_05)}
\end{equation}
where $A_{ex}^{(5)}$ is the fifth-order expansion of $A_{ex}\equiv A(t_{ex})$.
Generally, the complex tunneling velocity $v_{ex}$ can also be obtained by numerical solution of saddle point equations, and its real part $\mathrm{Re}[{v_{ex}}]$ can be considered as the exit velocity of the tunneling electron at the tunnel exit (or the initial velocity of the tunneling electron at the exit time $t_{ex}$) to describe its classical dynamics  in the laser field after tunneling.


\begin{figure}
	\begin{center}
		\includegraphics[width=1\columnwidth]{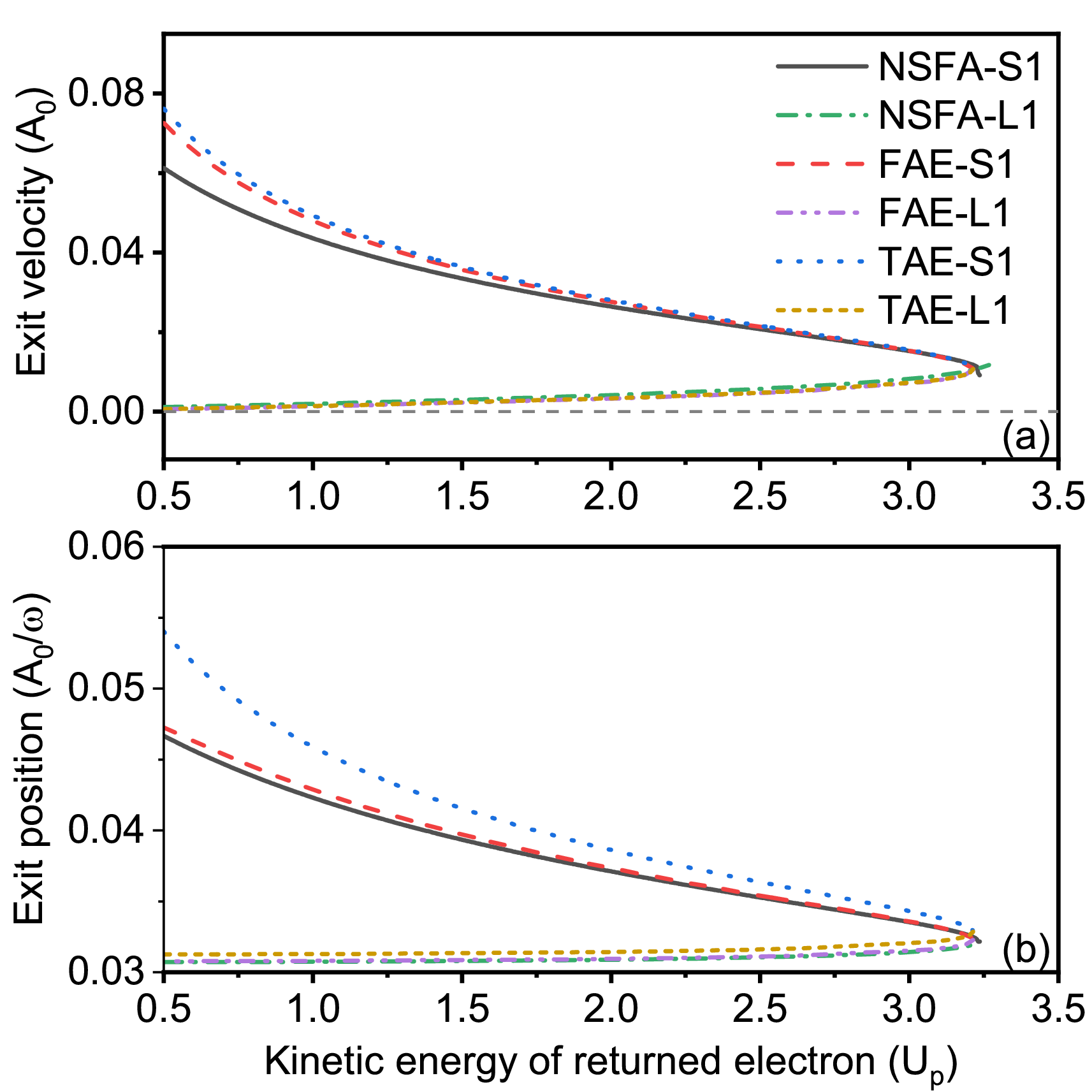}
	\end{center}
	\caption{The exit velocity $\mathrm{Re}[v_{ex}]$ (a) and the exit position $\mathrm{Re}[x_{ex}]$ (b) at the tunnel exit, as functions of the electron return kinetic energy, $E_{re}$, obtained using different methods for $\gamma=0.25$. The black solid and green dash-dotted lines represent the results of S1 and L1 trajectories obtained with NSFA, while the red long-dashed and purple dash-dot-dotted lines represent the results of S1 and L1 trajectories obtained with Eq. (\ref{eq:v(t_05)}) in (a) and Eq. (\ref{eq:x(t_0)5}) in (b). For comparison, the result of the third-order expansion are indicated using blue dotted and brown short-dashed lines in (a) and (b), respectively.}
	\label{fig:nonadiabatic}
\end{figure}

In adiabatic theory, the exit velocity is always equal to zero, and it can be used as a criterion to judge whether a method is adiabatic or not \cite{PhysRevA.98.013411}.
Here the zero-order expansion of $v_{ex}$ is equal to zero, which agrees with adiabatic approximation.
In Fig. \ref{fig:nonadiabatic}, we show the exit velocity and position obtained by different methods of NSFA and our analytical method. The result of zero-order expansion of $v_{ex}$ is indicated using a gray dashed line. It is obvious that the analytical results of the fifth-order expansion are far from $v_{ex}=0$ and vary with the kinetic energy of returned electron, aligning with the nonadiabatic approximation in terms of velocity criteria \cite{PhysRevA.98.013411}.
In particular, the analytical result of Eq. (\ref{eq:v(t_05)}) is in good agreement with the numerical one for the L1 trajectory. For the S1 trajectory, particularly in the low energy regions, there are somewhat differences between the prediction of Eq. (\ref{eq:v(t_05)}) and NSFA. As previously mentioned, the perturbative analytical approach may not be well-suited for this particular case.
This is because the effective adiabatic parameter $\gamma(t_{ex})$ for the S1 trajectory is larger than $\gamma$, and even greater than 1.

In the spirit of SFA theory, the complex position of the electron at the tunnel exit can be expressed as \cite{PhysRevA.98.013411}
\begin{equation}
	\begin{aligned}
	x_{ex}=\int_{t_i}^{t_{ex}} [ p_{s}-A(t)]  \,dt  \\		
	\end{aligned}
\label{eq:x(t_0)}
\end{equation}
Based on the analytical method, $t_i$ and $t_{ex}$ and $p_{s}$ all can be expressed as functions of $\tau_0$, so $x_{ex}$ can also be expanded according to the FAE method. That is

\begin{equation}
	\begin{aligned}
	x_{ex} \approx& x_{ex}^{(5)} \\
	\approx&  x_{ex0} +x_{ex2}\gamma^2 +x_{ex4}\gamma^4+i(x_{ex3}\gamma^3 +x_{ex5}\gamma^5), 		
	\end{aligned}
\label{eq:x(t_0)5}
\end{equation}
and here $x_{ex0}$ which represents the adiabatic approximation result is derived to be equal to 0. Similar to $v_{ex}$, the real part of the complex position $x_{ex}$ can be considered as the exit position of the tunneling electron at the tunnel exit (or the initial position at the exit time $t_{ex}$). In Fig.~\ref{fig:nonadiabatic} (b), the initial position obtained using the NSFA and FAE methods of Eq. (\ref{eq:x(t_0)5}) are compared. The analytical FAE results agree well with the numerical ones, especially for the L1 trajectory and in regions with higher photonic energy. These two quantities $\mathrm{Re}(v_{ex})$ and $\mathrm{Re}(x_{ex})$, which can be derived from our analytical method, provide accurate nonadiabatic initial conditions for developing analytical and quantitative Coulomb-corrected SFA models of HHG. Such models are important for constructing experimental procedures and deducing ultrafast time information from experimental measurements.

\section{Extended discussions\label{comparisons}}

Before conclusion, let us add more discussions on the significance of the present work and its correlation to other strong-field models.

\textit{Comparisons with previous work}. In the previous work in \cite{PhysRevA.106.053105}, the second-order expansion of the SFA formula for HHG with respect to the Keldysh parameter $\gamma$ is performed. With the second-order expansion, one can obtain the time-frequency information of the HHG electron trajectory but can not obtain the amplitude of the trajectory. In the present work, the third-order expansion and the fifth-order expansion of the SFA formula for HHG with respect to $\gamma$ are performed. With the third-order expansion, one can obtain the amplitude of the trajectory related to the adiabatic effect described by the ADK theory, while with the fifth-order expansion, one can obtain the amplitude of the trajectory related to the nonadiabatic effect beyond the description of ADK. Therefore, with these expansions, we can analytically obtain the coherent HHG spectrum and analytically study not only the time-frequency characteristic but also the coherent property of the HHG electron trajectories (including long and short trajectories of the first return and multiple returns). 

\textit{Application in Coulomb-corrected models}. With these high-order expansions, we can also accurately obtain the exit velocity and the exit position of the tunneling electron at the tunnel exit. These quantities are important in constructing a fully analytical and accurate HHG model. As it is well known that the SFA is not accurate enough due to the neglect of the important Coulomb effect, in recent years, trajectory-based Coulomb-included strong-field models such as MSFA \cite{Xie:20} have been developed to quantitatively describe the HHG. With using the exit velocity and position as the initial conditions weighted by the amplitude of the corresponding electron trajectory, these models consider the Coulomb effect through numerical solution of Newton equation including both electric-field force and Coulomb force after the electron tunnels out of the barrier.  On the other hand, in present ultrafast measurement experiments based on HHG, the time information is not directly measured. Instead, the experiments measure the coherent harmonic spectrum. By using a mapping between time and observables, the ultrafast dynamical information of the system can be deduced from the spectrum. To achieve an accurate retrieval of the time information, a precise and one-to-one mapping is needed. Such a mapping can be obtained from analytical and accurate SFA models. However, the trajectory-based Coulomb-included models mentioned above are semi-analytical and therefore can not provide such a mapping. The present work gives the analytical and accurate expressions for the initial conditions and the corresponding amplitude at the tunnel exit, providing a chance for constructing a fully analytical and accurate strong-field model for HHG. Since the ionization process is the first step of HHG, the perturbation expansion manner used in the present work can also be easily to generalized  to obtain a fully analytical model for above-threshold ionization \cite{PhysRevA.51.1495}.

\textit{Implication on HHG scaling law}. The scaling law for  wavelength-dependent HHG yields is important in shaping bright and short attosecond pulses. Through TDSE simulations, a scaling law of $\lambda^{-(5-6)}$ has been found \cite{PhysRevLett.98.013901}. This law can be mainly attributed to the spreading of the wavepacket of the rescattering electron, which is proportional to the excursion distance of the rescattering electron responsible for HHG. However, the influence of the ionization step in this law is not clear. To fully understand this law, an analytical SFA model for HHG is also needed. Indeed, the fully analytical method developed in the paper shows that the nonadiabatic effect related to the fifth-order expansion term depends on the laser wavelength and remarkably increases the HHG yields for cases of short laser wavelength, providing deeper insights into the wavelength-dependent scaling law of HHG yields. 

\textit{Application in propagation effect}. In actual experiments, the macroscopic propagation effect \cite{Weissenbilder2022}, which occurs when the harmonic goes through the media and is related to the phase matching, needs also be considered.  The theory calculation of this propagation effect is very time-consuming. As a result, the SFA is usually used in simulating the propagation. However, due to the need of solving the saddle-point equation for each spot of the laser beam, relevant simulations using the general SFA is also very time-consuming. In this case, the fully analytical method proposed in this paper, which does not need the numerical solution of saddle-point equation, will remarkably improve the calculation speed by at least an order of magnitude. It should be stressed that for some cases with small Keldysh parameters, the SFA can well reproduce the TDSE results, as shown in Fig. 9 in \cite{Le_2016}.

\begin{figure}
	\begin{center}
		\includegraphics[width=1\columnwidth]{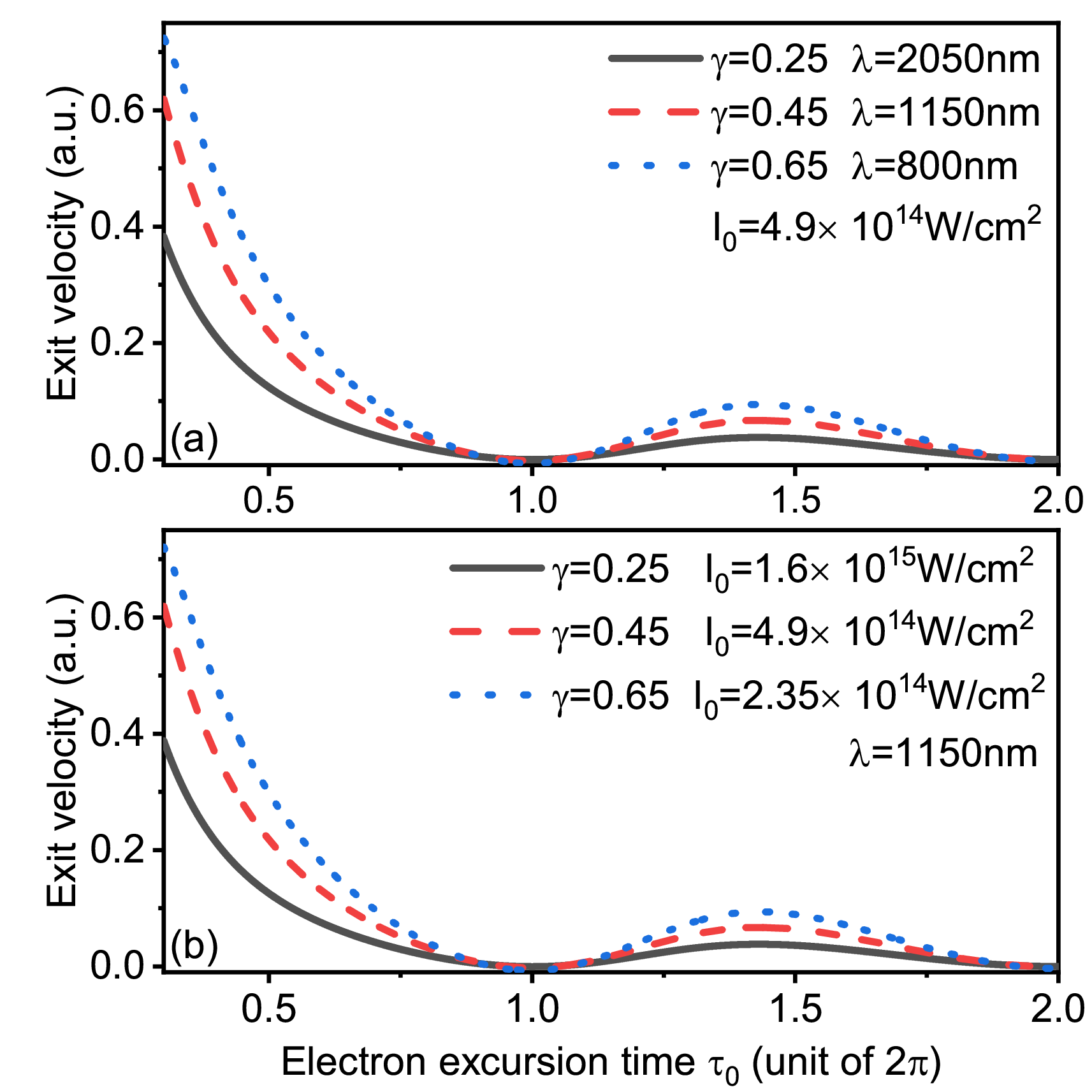}
	\end{center}
	\caption{
	Exit velocity Re$[v_{ex}]$ obtained with Eq. (\ref{eq:v(t_05)}) for the same laser and atomic parameters as those used in Fig~\ref{fig:theta5}.
	}
	\label{fig:vex_vs_tau0}
\end{figure}
\textit{Application to more complex cases}. For more complex atoms, according to the SFA, the saddle-point equation is unchanged in comparison with the simple ones, and the difficulty  exists in the accurate evaluation of the dipole transition matrix elements $d_r$ and $d_i$ in Eq.~(\ref{eq:frequency-dependent dipole moment2}). This difficulty can be overcome according to the QRS method \cite{PhysRevLett.100.013903}. For tailored fields, the formula obtained in the paper can not be directly used as they are obtained with the simple $\cos$-form laser field. However, the idea of perturbation expansion on the Keldysh parameter can still be used to analytically obtain the saddle points for laser fields having a more complex form. Specifically, to incorporate this method into tailored fields, one can use the slowly varying envelope approximation. Because the dipole moment employed in this method (as described in Eq.~(\ref{eq:frequency-dependent dipole moment1})) falls within one half of the optical cycle, we can, for tailored fields, treat the overall field as a combination of several half optical cycles \cite{Le_2016,PhysRevA.64.013409}. By calculating the results for each half optical cycle individually and subsequently adding these results together coherently, we can arrive at the final result.

\textit{Implication on nonadiabatic effects}. Generally, it is considered that for $\gamma \ll 1$, the tunneling ionization process is adiabatic \cite{Perelomov1966IONIZATIONOA}. The well-known ADK theory is considered to be applicable to  the case of $\gamma \ll 1$, thus describing the adiabatic tunneling process \cite{F_A_Ilkov_1992}. However, in usual experiments, the Keldysh parameter $\gamma$ is close to or larger than the unity, so nonadiabatic effects are expected to occur. The common manner for determining nonadiabatic effects is to evaluate the exit velocity $\mathrm{Re}[v_{ex}]$ at the tunnel exit. For $\mathrm{Re}[v_{ex}]\neq 0$, the tunneling process is considered nonadiabatic \cite{Perelomov1966IONIZATIONOA}.  In \cite{PhysRevLett.111.103003}, it is shown that for a large range of laser intensity, the experimental trends of the attoclock data support the use of adiabatic theory in describing tunneling ionization, in disagreement with the general opinion that the nonadiabatic theory needs to be used in tunneling ionization. In this paper, we explore the nonadiabatic effect on HHG where the ionization process is the first step. Indeed, our analytical results of perturbation expansion show that for each HHG electron trajectory, the nonadiabatic effect not only influences the exit velocity at the tunnel exit (see Eq.~(\ref{eq:v(t_05)})), but also affects the trajectory amplitude which is related to both the ionization and the recombination processes (see Eq. (\ref{eq:ADK})), suggesting that the nonadiabatic effect is fundamental in the SFA description of strong-field processes. We note that our results agree with the recent theory studies of attoclock \cite{chen2022responsetimephotoemissionquantumclassic,PhysRevA.107.043109,che2023} which show that the experimental data in \cite{PhysRevLett.111.103003} over a wide intensity range can be quantitatively reproduced through a developed SFA-trajectory-based nonadiabatic strong-field model that considers the Coulomb-related symmetry of the system in tunneling ionization \cite{PhysRevA.111.053118}. In particular, our results show that under certain laser parameters, even if the exit velocity is close to zero, the influence of nonadiabatic effects on trajectory amplitude still holds, and vice versa. These can be seen by comparing the amplitude results in Fig. {\ref{fig:theta5}} and the velocity results in Fig. {\ref{fig:vex_vs_tau0}}. The results in Fig. {\ref{fig:vex_vs_tau0}} are obtained with the same laser and atomic parameters as in Fig. {\ref{fig:theta5}}. One can observe that when a certain curve in Fig. {\ref{fig:theta5}} is near to the unity, the correspond one in Fig. {\ref{fig:vex_vs_tau0}} can not be near to zero, and vice versa. Therefore, we consider that nonadiabatic effects usually exist in tunneling-triggered strong-field processes, reflecting the essential dynamical properties of laser-driven quantum systems.

\textit{Limitations of the present method}. The perturbation expansion method used in the paper is based on SFA, so the applicability of the method is similar to the SFA, which implies that this method can only be used for cases of strong low-frequency laser fields. In addition, as this method is related to the perturbation expansion on the Keldysh parameter $\gamma$ up to the fifth order, it is more applicable for smaller $\gamma$ with $\gamma\ll1 $. In fact, our simulations show that this method works well for cases of  $\gamma < 0.65$, as shown in Fig. {\ref{fig:theta5}}. For these cases, the solutions of this method are consistently stable and are in quantitative agreement with the results obtained through numerical solutions of the SFA saddle-point equation.

\section{Conclusion\label{concl}}
In summary, in this paper, we propose a fully analytical method that provides a detailed description of HHG. This method is based on SFA and combines the saddle point method with perturbation expansion. By expanding the ionization time, return time, and canonical momentum with respect to the Keldysh parameter, almost every physical quantity can be expressed as a function of the independent variable $\tau_0$, including the HHG yield. Therefore, this method offers a completely analytical alternative to the commonly used numerical SFA method for HHG.

By comparing the HHG yields of different electron trajectories obtained by the perturbative analytical method and the numerical saddle point method, we demonstrate that the perturbation analytical method is reliable for cases of $\gamma\leqslant 0.27$ with a third-order expansion and $\gamma\leqslant 0.65$ with a fifth-order expansion. Moreover, the accuracy of the analytical method is further confirmed when considering the influence of the interference of different quantum trajectories. We demonstrate that the TAE term of the HHG yield provides a good approximation for the L1 trajectory, which corresponds to a small effective adiabatic parameter. However, it is less accurate for the S1 trajectory, which corresponds to a large effective adiabatic parameter. Including the fifth-order expansion term results in good agreement with the numerical results.


Furthermore, we show the important influence of high-order expansion terms on the wavelength-dependent HHG yield. Our formulas indicate that the third-order expansion term, which resembles an ADK-like form, depends solely on the binding energy and laser intensity. In contrast, the fifth-order expansion term also takes the laser wavelength into account. As the third-order expansion reflects the adiabatic effect in tunneling ionization, the fifth-order expansion contains nonadiabatic effects and can remarkably increase the HHG yields for cases of shorter laser wavelengths. The wavelength-dependent the fifth-order contribution also provides deeper insights into the wavelength-dependent HHG yields which are important in shaping bight attosecond pulses in experiments.


Additionally, the fifth-order exit velocity and position at the tunnel exit are shown to be nonadiabatic, capturing the nonadiabatic electron dynamics under the barrier. Considering nonadiabatic effects in the exit velocity and position allows for a more accurate description of electron dynamics, resulting in more precise HHG spectra. The exit velocity and position of the tunneling electron at the tunnel exit influence importantly on the subsequent motion of the electron, and can be used as the initial conditions for trajectory-based Coulomb-corrected SFA models. The analytical description of these initial conditions provides a chance for constructing a fully analytical and quantitative Coulomb-corrected SFA model of HHG.

In actual experiments, the macroscopic propagation effect mainly associated with phase matching has also an important influence on HHG. Theory simulations of macroscopic propagation of HHG are very time-consuming, even the SFA is used. With the fully-analytical SFA method developed here instead of the numerical procedure in the general SFA, the calculation time of macroscopic propagation can be reduced by at least one order of magnitude, facilitating related studies.

\begin{acknowledgments}
This work was supported by the National Natural Science Foundation of China (Grant No. 12374322 and No. 12104501), the National Key Research and Development Program of China (Grant No. 2022YFE0111500), and NSF Investigator-Initiated Research (Grant No. 2208040). Thanks to Computing Center in Xi'an for the computing power support .
\end{acknowledgments}

\section*{Data Availability}
The data that support the findings of this article are openly available at \cite{sun_2026_18288498}.

\appendix
\section{The scaling process of saddle point equation}
Based on Eq. (\ref{eq:ps(t)}), Eq. (\ref{eq:saddle point equationb}) can be rewritten in a more specific form:
\begin{equation}
	\begin{aligned}
		\frac{A_0^2}{2} [\frac{\cos (t_r)-\cos (t_r-\tau )}{\tau }+\sin(t_r-\tau)]^2=-I_p.
	\end{aligned}
	\label{eq:saddle point equationb1}
\end{equation}
When scaled by $U_p$, the equation above can be simplified to the form related to the adiabatic parameter $\gamma$:
\begin{equation}
	\begin{aligned}
		{\frac{1}{\tau}[\cos(t_r)-\cos(t_r-\tau)]+\sin(t_r-\tau)}= \pm i\gamma.
	\end{aligned}
	\label{eq:saddle point equationb2}
\end{equation}
Here, $t_r$ can be written as $(t_r - \tau/2) + \tau/2$ and $t_r - \tau$ can be written as $(t_r - \tau/2) - \tau/2$. In the following, as an example, we discuss the case of positive sign in the above expression. With using trigonometric functions, one can obtain the following equation:
\begin{equation}
	\begin{aligned}
		\sin(t_r-\tau/2&)[\cos(\tau/2)-\frac{2\sin(\tau/2)}{\tau}]\\
		&-\cos(t_r-\tau/2)\sin(\tau/2)= i\gamma.
	\end{aligned}
	\label{eq:saddle point equationb3}
\end{equation}
Following the definitions $a(\tau)=\cos(\tau/2)-2\sin(\tau/2)/\tau$ and $s(\tau)= \sin(\tau/2)$, Eq. (\ref{eq:saddle point equationb3}) can be written in the form of Eq. (\ref{eq:saddleb reduced form}). 

\section{The derivation of the kinetic energy and return time of returned electron}
As indicated in Eq. (\ref{eq:saddle point equationb}), the first term on the left side of the equation represents the electron velocity at the moment of ionization $t_r$, while the second term represents the electron velocity at the moment $t_r-\tau$. Referring to Eq. (\ref{eq:ps(t)}), one can determine the electron velocity at the return moment $t_r$:
\begin{equation}
	\begin{aligned}
	v(t_r) = &p_s-A(t_r) \\=&A_0 [\frac{\cos (t_r)-\cos (t_r-\tau )}{\tau }+\sin (t_r)].
	\end{aligned}
	\label{eq:v(t_r)}
\end{equation}
By using the definition of $a(\tau)$ and $s(\tau)$, the kinetic energy of the returned electron can be expressed as
\begin{equation}
	E_{re}=2U_p
		[\sin\left( t_r-\frac {\tau}{2}\right) a\left( \tau \right) +\cos\left( t_r-\frac {\tau}{2}\right) s\left( \tau\right)]^2.
	\label{eq:return kinetic energy1}
\end{equation}

By utilizing the trigonometric identity $\sin\left( t_r-\tau/2\right)^2+\cos\left( t_r-\tau/2\right)^2=1$ in Eq. (\ref{eq:saddleb reduced form}), we can derive the the following formula
\begin{equation}
	\sin(t_r-\frac{\tau}{2})=\frac{i\gamma a\left( \tau \right) + s\left( \tau \right) \sqrt{a^2\left( \tau \right) +s^2\left( \tau \right) +\gamma^2}}{ a^2\left( \tau \right) +s^2\left( \tau \right)}, 
	\label{eq:sin}
\end{equation} 
\begin{equation}
	\cos(t_r-\frac{\tau}{2})=\frac{-i\gamma s\left( \tau \right) + a\left( \tau \right) \sqrt{s^2\left( \tau \right) +a^2\left( \tau \right) +\gamma^2}}{ s^2\left( \tau \right) +a^2\left( \tau \right)}.
	\label{eq:cos}
\end{equation}
By substituting Eqs. (\ref{eq:sin}) and Eq. (\ref{eq:cos}) into Eq. (\ref{eq:return kinetic energy1}), one can derive the kinetic energy of the returned electron as a function of $\tau$, expressed as Eq. (\ref{eq:kinetic energy}).
Then the return time can be expressed as
\begin{equation}
	t_r=\frac{\tau}{2}+\arctan \frac{\sin(t_r-\frac{\tau}{2})}{\cos(t_r-\frac{\tau}{2})} .
	\label{eq:arctan}
\end{equation}
Inserting Eq. (\ref{eq:sin}) and Eq. (\ref{eq:cos}) into Eq. (\ref{eq:arctan}), one can obtain the following expression for the electron return time 
\begin{equation}
	\begin{aligned}
	&t_r=\frac{\tau }{2}+ \\ &\arctan(\frac{a \sqrt{a^2+\gamma ^2+s^2}-is \gamma }{a^2+s^2}, \frac{s \sqrt{a^2+\gamma ^2+s^2}+i a \gamma }{a^2+s^2}).
	\end{aligned}	
	\label{eq:return time}
\end{equation}

\section{The derivation of $\theta_5$}
According to Eqs. (\ref{eq:theta}) and (\ref{eq:Theta}), $\theta$ is composed of two parts: the scaled quasiclassical action and the product of photon energy and return time. By scaling $S(p_{s}, t_r, \tau) = (U_p/\omega) \alpha$ and $\Omega t_{r} / \omega = (U_p/\omega) \beta$, $\theta$ can be expressed as
\begin{equation}
	\begin{aligned}
\theta&=\alpha+\beta. \\
	\end{aligned}
\end{equation}
Therefore, the FAE term of $\theta$ can be expressed as
\begin{equation}
	\begin{aligned}
\theta_5&=\alpha_5+\beta_5, \\
\label{eq:theta_5}
	\end{aligned}
\end{equation}
where $\alpha_5$ and $\beta_5$ are the fifth-order terms of $\alpha$ and $\beta$, respectively. Starting from $\beta$, due to the relationship $\Omega=E_{re}+I_p$, $\beta$ can be expressed as
\begin{equation}
	\begin{aligned}
	\beta&=\frac{(E_{re}+I_p)}{U_p}t_r \\
	\end{aligned}
	\label{eq:beta}
\end{equation}
Inserting Eqs.(\ref{eq: third order kinetic energy}) and (\ref{eq:tr}) into Eq.(\ref{eq:beta}), and keeping terms up to $\gamma^5$, one can obtain
\begin{equation}
	\begin{aligned}
	\beta_5&=(E_{re2}+2)t_{r3}. \\
	\end{aligned}
	\label{eq:beta_5}
\end{equation}
As for $\alpha$, using the definition $E(t)=E_0 \cos(t)$, Eq. (\ref{eq:ps(t)}) can be used to replaced $\textbf{p}$ in Eq. (\ref{eq:quasiclassical action.}). Then $S(p_{st}, t_r, \tau)$ can be expressed as
\begin{equation}
	\begin{aligned}
	S=\frac{1}{\omega} \int_{t_i}^{t_r}{\left[\frac{A_0^2\left( \frac{\cos\left(t_r\right)-\cos\left(t_i\right)}{\tau}-\sin\left(t\right)    \right) ^2}{2}+I_p\right] dt }. 
	\end{aligned}
	\label{eq:SS}
\end{equation}
Scaling $S(p_{st}, t_r, \tau)$ with $(U_p/\omega)$, one can further obtain the expression of $\alpha$. That is 
\begin{equation}
	\begin{aligned}
	\alpha= 2\int_{t_i}^{t_r}{\left[\left( \frac{\cos\left(t_r\right)-\cos\left(t_i\right)}{\tau}-\sin\left(t\right)    \right) ^2+\gamma^2\right] dt }. 
	\end{aligned}
	\label{eq:alpha1}
\end{equation}
Solving integral Eq. (\ref{eq:alpha1}), $\alpha$ can be rewritten as
\begin{equation}
	\begin{aligned}
		\alpha=&2 \gamma ^2 \tau + \frac{-2+2 \cos \tau +\tau^2 }{\tau}+\\
		&\frac{-\cos 2t_i +(2-\tau \sin \tau ) \cos (t_r+t_i )-\cos 2 t_r}{\tau }.
	\label{eq:alpha2}
	\end{aligned}
\end{equation}
Based on the FAE method, $\alpha_5$ can be expressed as 
\begin{equation}
	\begin{aligned}
		\alpha_5=\alpha_A+\alpha_B+\alpha_C+\alpha_D+\alpha_E+\alpha_F+\alpha_G.
	\label{eq:alpha_5}
	\end{aligned}
\end{equation}
Here $\alpha_{A-G}$ are
\onecolumngrid
\begin{equation}
	\begin{aligned}
	\alpha_A=&2\tau_3+2\tau_1 t_{r2}^2\cos2(t_{i0}) +2\tau_1^2t_{r3}\cos(2t_{i0})-\frac{2}{15} \tau_1^5\cos(2t_{i0})-2\tau_1^2\tau_3\cos(2t_{i0})+2t_{r2}t_{r3}\sin(2t_{r0})-2t_{r2}t_{r3}\sin(2t_{i0})\\ 
	&+ \frac{4}{3}\tau_1^3t_{r2}\sin(2t_{i0})+2\tau_3t_{r2}\sin(2t_{i0}), 
	\end{aligned}
	\label{eq:alpha_A}
\end{equation}
\begin{equation}
	\begin{aligned}
	\alpha_B=&\frac{1}{\tau_0}\times [-\frac{1}{60} \tau_1^5 \sin (\tau_0)-\tau_1^2 \tau_3 \sin (\tau_0)-\frac{4}{15} \tau_1^5 \sin (2t_{i0})+\frac{1}{60} \tau_1^5 \sin (t_{r0}+t_{i0})-4 \tau_1^2 \tau_3 \sin (2t_{i0})+\tau_1^2 \tau_3 \sin (t_{r0}+t_{i0})\\
	&+4 \tau_1 t_{r2}^2 \sin (2t_{i0})-4 \tau_1 t_{r2}^2 \sin (t_{r0}+t_{i0})+\frac{2}{3} \tau_1^3 t_{r2} \cos (t_{r0}+t_{i0})+4 \tau_3 t_{r2} \cos (t_{r0}+t_{i0})+4 t_{r2} t_{r3} \cos (2 t_{r0})\\
	&-8 t_{r2} t_{r3} \cos (t_{r0}+t_{i0})+4 \tau_1^2 t_{r3} \sin (2t_{i0})-2 \tau_1^2 t_{r3} \sin (t_{r0}+t_{i0})]+\frac{4}{3} t_{r2} \cos (2t_{i0}) \left(-2 \tau_1^3-3 \tau_3+3 t_{r3}\right), 
	\end{aligned}
	\label{eq:alpha_B}
\end{equation}
\begin{equation}
	\begin{aligned}
	\alpha_C=&-\frac{1}{12\tau_0^2}\times [24\tau_1t_{r2}^2\cos(2t_{r0})-8\tau_1(\tau_1^4+9\tau_1\tau_3-6\tau_1t_{r3}-3t_{r2}^2)\cos(2t_{i0})-48\tau_1t_{r2}^2\cos(t_{r0}+t_{i0})+\tau_1^5\cos(t_{r0}+t_{i0})\\
	&-48\tau_1^2t_{r3}\cos(t_{r0}+t_{i0})+36\tau_1^2\tau_3\cos(t_{r0}+t_{i0})+\tau_1^5\cos(\tau_0)+36\tau_1^2\tau_3\cos(\tau_0)+24\tau_3t_{r2}\sin(2t_{r0})\\
	&+48\tau_1^3t_{r2}\sin(2t_{i0})+24\tau_3t_{r2}\sin(2t_{i0})-24(\tau_1^3+2\tau_3)t_{r2}\sin(t_{r0}+t_{i0})], 
	\end{aligned}
	\label{eq:alpha_C}
\end{equation}
\begin{equation}
	\begin{aligned}
	\alpha_D=&\frac{1}{3 \tau_0^3}\times 2 \tau_1^2[\left(2 \tau_1^3+9 \tau_3\right) \sin (t_{r0}+t_{i0})-\left(\tau_1^3+18 \tau_3\right) \cos ( t_{r0}) \sin ( t_{r0}+t_{i0})+6 \tau_1 t_{r2} \cos (2t_{i0} )-6  \tau_1 t_{r2} \cos (t_{r0}+ t_{i0})\\ &+
	12 t_{r3} \sin ^2\left(\frac{ { \tau_0}}{2}\right) \sin (t_{r0}+t_{i0})], 
	\end{aligned}
	\label{eq:alpha_D}
\end{equation}
\begin{equation}
	\begin{aligned}
	\alpha_E=&-\frac{1}{3 \tau_0^4} \times \tau_1^2[6 \tau_3-2 \left(\tau_1^3+6 \tau_3\right) \cos (t_{r0}) \cos (t_{i0})+\left(2 \tau_1^3+3 \tau_3\right) \cos (t_{r0}+t_{i0})+8 \tau_1 t_{r2} \sin ^2\left(\frac{\tau_0}{2}\right) \sin (t_{r0}+t_{i0})\\
	&+3 \tau_3 \cos (2 t_{r0})], 
	\end{aligned}
	\label{eq:alpha_E}
\end{equation}
\twocolumngrid
\begin{equation}
	\begin{aligned}
	\alpha_F=&-\frac{2\tau_1^5[\sin (t_{r0}+t_{i0})-2 \cos (t_{r0}) \sin (t_{i0})]}{ \tau_0^5} , 
	\end{aligned}
	\label{eq:alpha_F}
\end{equation}
\begin{equation}
	\begin{aligned}
	\alpha_G=&\frac{ 8\tau_1^5[\sin ^2\left(\frac{{\tau_0}}{2}\right) \sin ^2\left(t_{r0}-\frac{{\tau_0}}{2}\right)]}{ \tau_0^6}.
	\end{aligned}
	\label{eq:alpha_G}
\end{equation}
By substituting Eq. (\ref{eq:beta_5}) and Eq. (\ref{eq:alpha_5}) into Eq. (\ref{eq:theta_5}), one can obtain $\theta_5$ in relation to ionization time, return time, and excursion time.
\bibliography{refs1.bib}

@article{10.1063/1.3069511,
  author   = {Chen, Y. J. and Liu, J. and Hu, Bambi},
  title    = {{Reading molecular messages from high-order harmonic spectra at different orientation angles}},
  journal  = {J. Chem. Phys},
  volume   = {130},
  number   = {4},
  pages    = {044311},
  year     = {2009},
  month    = {01},
  issn     = {0021-9606},
  doi      = {10.1063/1.3069511},
  url      = {https://doi.org/10.1063/1.3069511}
}

@article{1986Tunnel,
  title   = {Tunnel ionization of complex atoms and of atomic ions in an alternating electromagnetic field},
  author  = { Ammosov, Maxim V.  and  Delone, Nikolai B.  and  Krainov, Vladimir P. },
  journal = {Sov. Phys. JETP},
  volume  = {64},
  pages   = {1191},
  year    = {1986}
}

@article{Delone:91,
  author    = {N. B. Delone and V. P. Krainov},
  journal   = {J. Opt. Soc. Am. B},
  number    = {6},
  pages     = {1207--1211},
  publisher = {Optica Publishing Group},
  title     = {Energy and angular electron spectra for the tunnel ionization of atoms by strong low-frequency radiation},
  volume    = {8},
  month     = {Jun},
  year      = {1991},
  url       = {https://opg.optica.org/josab/abstract.cfm?URI=josab-8-6-1207},
  doi       = {10.1364/JOSAB.8.001207}
}

@article{doi:10.1080/09500340412331301542,
  author    = {K. Varjú and Y. Mairesse and B. Carré and M. B. Gaarde and P. Johnsson and S. Kazamias and R. López-Martens and J. Mauritsson and K. J. Schafer and PH. Balcou and A. L'Huillier and P. Salières},
  title     = {Frequency chirp of harmonic and attosecond pulses},
  journal   = {J. Mod. Opt.},
  volume    = {52},
  number    = {2-3},
  pages     = {379--394},
  year      = {2005},
  publisher = {Taylor \& Francis},
  doi       = {10.1080/09500340412331301542}
}

@article{doi:10.1126/sciadv.aao5207,
  author  = {Nicolas Tancogne-Dejean  and Angel Rubio },
  title   = {Atomic-like high-harmonic generation from two-dimensional materials},
  journal = {Sci. Adv.},
  volume  = {4},
  number  = {2},
  pages   = {eaao5207},
  year    = {2018},
  doi     = {10.1126/sciadv.aao5207},
  url     = {https://www.science.org/doi/abs/10.1126/sciadv.aao5207}
}

@article{doi:10.1126/science.1059413,
  author   = {P. M. Paul and E. S. Toma and P. Breger and G. Mullot and F. Augé and Ph. Balcou and H. G. Muller and P. Agostini},
  title    = {Observation of a Train of Attosecond Pulses from High Harmonic Generation},
  journal  = {Science},
  volume   = {292},
  number   = {5522},
  pages    = {1689-1692},
  year     = {2001},
  doi      = {10.1126/science.1059413},
  url      = {https://www.science.org/doi/abs/10.1126/science.1059413}
}

@article{doi:10.1126/science.aac9755,
  author  = {Dimitar Popmintchev and Carlos Hernández-García  and Franklin Dollar  and Christopher Mancuso  and Jose A. Pérez-Hernández  and Ming-Chang Chen  and Amelia Hankla  and Xiaohui Gao  and Bonggu Shim  and Alexander L. Gaeta  and Maryam Tarazkar  and Dmitri A. Romanov  and Robert J. Levis  and Jim A. Gaffney  and Mark Foord  and Stephen B. Libby  and Agnieszka Jaron-Becker  and Andreas Becker  and Luis Plaja  and Margaret M. Murnane  and Henry C. Kapteyn  and Tenio Popmintchev },
  title   = {Ultraviolet surprise: Efficient soft x-ray high-harmonic generation in multiply ionized plasmas},
  journal = {Science},
  volume  = {350},
  number  = {6265},
  pages   = {1225-1231},
  year    = {2015},
  doi     = {10.1126/science.aac9755},
  url     = {https://www.science.org/doi/abs/10.1126/science.aac9755}
}

@article{doi:10.34133/2022/9767251,
  author  = {Mengxue Guan  and Daqiang Chen  and Shiqi Hu  and Hui Zhao  and Peiwei You  and Sheng Meng },
  title   = {Theoretical Insights into Ultrafast Dynamics in Quantum Materials},
  journal = {Ultrafast Science},
  volume  = {2022},
  number  = {9767251},
  pages   = {9767251},
  year    = {2022},
  doi     = {10.34133/2022/9767251},
  url     = {https://spj.science.org/doi/abs/10.34133/2022/9767251}
}

@article{doi:10.34133/ultrafastscience.0034,
  author  = {Chunmei Zhang  and Graham Brown  and Dong Hyuk Ko  and P. B. Corkum },
  title   = {Optical Measurement of Photorecombination Time Delays},
  journal = {Ultrafast Science},
  volume  = {3},
  number  = {},
  pages   = {0034},
  year    = {2023},
  doi     = {10.34133/ultrafastscience.0034},
  url     = {https://spj.science.org/doi/abs/10.34133/ultrafastscience.0034}
}

@article{F_A_Ilkov_1992,
  doi       = {10.1088/0953-4075/25/19/011},
  url       = {https://dx.doi.org/10.1088/0953-4075/25/19/011},
  year      = {1992},
  month     = {oct},
  publisher = {},
  volume    = {25},
  number    = {19},
  pages     = {4005},
  author    = {F A Ilkov and J E Decker and S L Chin},
  title     = {Ionization of atoms in the tunnelling regime with experimental evidence using Hg atoms},
  journal   = {J. Phys. B}
}

@article{FARKAS1992447,
  title    = {Proposal for attosecond light pulse generation using laser induced multiple-harmonic conversion processes in rare gases},
  journal  = {Phys. Lett. A},
  volume   = {168},
  number   = {5},
  pages    = {447-450},
  year     = {1992},
  issn     = {0375-9601},
  doi      = {https://doi.org/10.1016/0375-9601(92)90534-S},
  url      = {https://www.sciencedirect.com/science/article/pii/037596019290534S},
  author   = {Gy. Farkas and Cs. Tóth}
}

@article{Ferray_1988,
  doi       = {10.1088/0953-4075/21/3/001},
  url       = {https://dx.doi.org/10.1088/0953-4075/21/3/001},
  year      = {1988},
  month     = {feb},
  publisher = {},
  volume    = {21},
  number    = {3},
  pages     = {L31},
  author    = {M Ferray and A L'Huillier and X F Li and L A Lompre and G Mainfray and C Manus},
  title     = {Multiple-harmonic conversion of 1064 nm radiation in rare gases},
  journal   = {J. Phys. B}
}

@article{Hentschel2001,
  author   = {Hentschel,M.
              and Kienberger, R.
              and Spielmann, Ch.
              and Reider, G. A.
              and Milosevic, N.
              and Brabec, T.
              and Corkum, P.
              and Heinzmann, U.
              and Drescher, M.
              and Krausz, F.},
  title    = {Attosecond metrology},
  journal  = {Nature},
  year     = {2001},
  month    = {Nov},
  day      = {01},
  volume   = {414},
  number   = {6863},
  pages    = {509-513},
  issn     = {1476-4687},
  doi      = {10.1038/35107000},
  url      = {https://doi.org/10.1038/35107000}
}

@article{Keldysh65,
  title   = {Ionization in the Field of a Strong Electromagnetic Wave},
  volume  = {20},
  pages   = {1307--1314},
  number  = {5},
  journal = {Sov. Phys. JETP},
  author  = {Keldysh, L.V.},
  year    = {1965}
}

@article{Le_2016,
  doi       = {10.1088/0953-4075/49/5/053001},
  url       = {https://dx.doi.org/10.1088/0953-4075/49/5/053001},
  year      = {2016},
  month     = {feb},
  publisher = {IOP Publishing},
  volume    = {49},
  number    = {5},
  pages     = {053001},
  author    = {Anh-Thu Le and Hui Wei and Cheng Jin and C D Lin},
  title     = {Strong-field approximation and its extension for high-order harmonic generation with mid-infrared lasers},
  journal   = {J. Phys. B}
}

@article{Lewenstein94,
  title    = {Theory of high-harmonic generation by low-frequency laser fields},
  volume   = {49},
  issnx    = {1050-2947, 1094-1622},
  urlx     = {https://link.aps.org/doi/10.1103/PhysRevA.49.2117},
  doixx    = {10.1103/PhysRevA.49.2117},
  pages    = {2117--2132},
  number   = {3},
  journal  = {Phys. Rev. A},
  author   = {Lewenstein, M. and Balcou, Ph. and Ivanov, M. Yu. and L’Huillier, Anne and Corkum, P. B.},
  urlxyear = {2021-04-08},
  year     = {1994},
  langid   = {english}
}

@article{McPherson:87,
  author    = {A. McPherson and G. Gibson and H. Jara and U. Johann and T. S. Luk and I. A. McIntyre and K. Boyer and C. K. Rhodes},
  journal   = {J. Opt. Soc. Am. B},
  number    = {4},
  pages     = {595--601},
  publisher = {Optica Publishing Group},
  title     = {Studies of multiphoton production of vacuum-ultraviolet radiation in the rare gases},
  volume    = {4},
  month     = {Apr},
  year      = {1987},
  url       = {https://opg.optica.org/josab/abstract.cfm?URI=josab-4-4-595},
  doi       = {10.1364/JOSAB.4.000595}
}

@article{Perelomov1966IONIZATIONOA,
  title   = {IONIZATION OF ATOMS IN AN ALTERNATING ELECTRIC FIELD},
  author  = {Askold M. Perelomov and Valentin S. Popov and M. V. Terentev},
  journal = {Sov. Phys. JETP},
  year    = {1966},
  volume  = {23},
  pages   = {924},
  url     = {https://api.semanticscholar.org/CorpusID:118044175}
}

@article{PhysRevA.106.053105,
  title     = {Influence of initial tunneling step on the return energy of high-order harmonic generation},
  author    = {Gao, Xu-Zhen and Landsman, Alexandra S. and Cao, Huabao and Zhang, Yanpeng and Wang, Yishan and Fu, Yuxi and Pi, Liang-Wen},
  journal   = {Phys. Rev. A},
  volume    = {106},
  issue     = {5},
  pages     = {053105},
  numpages  = {10},
  year      = {2022},
  month     = {Nov},
  publisher = {American Physical Society},
  doi       = {10.1103/PhysRevA.106.053105},
  url       = {https://link.aps.org/doi/10.1103/PhysRevA.106.053105}
}

@article{PhysRevA.39.5751,
  title     = {Multiple-harmonic generation in rare gases at high laser intensity},
  author    = {Li, X. F. and L'Huillier, A. and Ferray, M. and Lompr\'e, L. A. and Mainfray, G.},
  journal   = {Phys. Rev. A},
  volume    = {39},
  issue     = {11},
  pages     = {5751--5761},
  numpages  = {0},
  year      = {1989},
  month     = {Jun},
  publisher = {American Physical Society},
  doi       = {10.1103/PhysRevA.39.5751},
  url       = {https://link.aps.org/doi/10.1103/PhysRevA.39.5751}
}

@article{PhysRevA.64.013409,
  title     = {Nonadiabatic tunnel ionization: Looking inside a laser cycle},
  author    = {Yudin, Gennady L. and Ivanov, Misha Yu.},
  journal   = {Phys. Rev. A},
  volume    = {64},
  issue     = {1},
  pages     = {013409},
  numpages  = {4},
  year      = {2001},
  month     = {Jun},
  publisher = {American Physical Society},
  doi       = {10.1103/PhysRevA.64.013409},
  url       = {https://link.aps.org/doi/10.1103/PhysRevA.64.013409}
}

@article{PhysRevA.78.053414,
  title     = {Probing $\mathrm{H}_{2}{}^{+}$ vibrational motions with high-order harmonic generation},
  author    = {Zhao, Jing and Zhao, Zengxiu},
  journal   = {Phys. Rev. A},
  volume    = {78},
  issue     = {5},
  pages     = {053414},
  numpages  = {5},
  year      = {2008},
  month     = {Nov},
  publisher = {American Physical Society},
  doi       = {10.1103/PhysRevA.78.053414},
  url       = {https://link.aps.org/doi/10.1103/PhysRevA.78.053414}
}

@article{PhysRevA.87.063811,
  title     = {High-order-harmonic generation from dense water microdroplets},
  author    = {Kurz, Heiko G. and Steingrube, Daniel S. and Ristau, Detlev and Lein, Manfred and Morgner, Uwe and Kova\ifmmode \check{c}\else \v{c}\fi{}ev, Milutin},
  journal   = {Phys. Rev. A},
  volume    = {87},
  issue     = {6},
  pages     = {063811},
  numpages  = {9},
  year      = {2013},
  month     = {Jun},
  publisher = {American Physical Society},
  doi       = {10.1103/PhysRevA.87.063811},
  url       = {https://link.aps.org/doi/10.1103/PhysRevA.87.063811}
}

@article{PhysRevA.96.023406,
  title     = {Adiabatic-limit Coulomb factors for photoelectron and high-order-harmonic spectra},
  author    = {Frolov, M. V. and Manakov, N. L. and Minina, A. A. and Popruzhenko, S. V. and Starace, Anthony F.},
  journal   = {Phys. Rev. A},
  volume    = {96},
  issue     = {2},
  pages     = {023406},
  numpages  = {12},
  year      = {2017},
  month     = {Aug},
  publisher = {American Physical Society},
  doi       = {10.1103/PhysRevA.96.023406},
  url       = {https://link.aps.org/doi/10.1103/PhysRevA.96.023406}
}

@article{PhysRevA.98.013411,
  title     = {Tunneling criteria and a nonadiabatic term for strong-field ionization},
  author    = {Ni, Hongcheng and Eicke, Nicolas and Ruiz, Camilo and Cai, Jun and Oppermann, Florian and Shvetsov-Shilovski, Nikolay I. and Pi, Liang-Wen},
  journal   = {Phys. Rev. A},
  volume    = {98},
  issue     = {1},
  pages     = {013411},
  numpages  = {13},
  year      = {2018},
  month     = {Jul},
  publisher = {American Physical Society},
  doi       = {10.1103/PhysRevA.98.013411},
  url       = {https://link.aps.org/doi/10.1103/PhysRevA.98.013411}
}

@article{PhysRevLett.100.013903,
  title     = {Accurate Retrieval of Structural Information from Laser-Induced Photoelectron and High-Order Harmonic Spectra by Few-Cycle Laser Pulses},
  author    = {Morishita, Toru and Le, Anh-Thu and Chen, Zhangjin and Lin, C. D.},
  journal   = {Phys. Rev. Lett.},
  volume    = {100},
  issue     = {1},
  pages     = {013903},
  numpages  = {4},
  year      = {2008},
  month     = {Jan},
  publisher = {American Physical Society},
  doi       = {10.1103/PhysRevLett.100.013903},
  url       = {https://link.aps.org/doi/10.1103/PhysRevLett.100.013903}
}

@article{PhysRevLett.111.103003,
  title     = {Probing Nonadiabatic Effects in Strong-Field Tunnel Ionization},
  author    = {Boge, R. and Cirelli, C. and Landsman, A. S. and Heuser, S. and Ludwig, A. and Maurer, J. and Weger, M. and Gallmann, L. and Keller, U.},
  journal   = {Phys. Rev. Lett.},
  volume    = {111},
  issue     = {10},
  pages     = {103003},
  numpages  = {5},
  year      = {2013},
  month     = {Sep},
  publisher = {American Physical Society},
  doi       = {10.1103/PhysRevLett.111.103003},
  url       = {https://link.aps.org/doi/10.1103/PhysRevLett.111.103003}
}

@article{PhysRevLett.124.203901,
  title     = {Impact of Statistical Fluctuations on High Harmonic Generation in Liquids},
  author    = {Zeng, Ai-Wu and Bian, Xue-Bin},
  journal   = {Phys. Rev. Lett.},
  volume    = {124},
  issue     = {20},
  pages     = {203901},
  numpages  = {5},
  year      = {2020},
  month     = {May},
  publisher = {American Physical Society},
  doi       = {10.1103/PhysRevLett.124.203901},
  url       = {https://link.aps.org/doi/10.1103/PhysRevLett.124.203901}
}

@article{PhysRevLett.129.167402,
  title     = {Proposal for High-Energy Cutoff Extension of Optical Harmonics of Solid Materials Using the Example of a One-Dimensional {ZnO} Crystal},
  author    = {Lang, Yue and Peng, Zhaoyang and Liu, Jinlei and Zhao, Zengxiu and Ghimire, Shambhu},
  journal   = {Phys. Rev. Lett.},
  volume    = {129},
  issue     = {16},
  pages     = {167402},
  numpages  = {6},
  year      = {2022},
  month     = {Oct},
  publisher = {American Physical Society},
  doi       = {10.1103/PhysRevLett.129.167402},
  url       = {https://link.aps.org/doi/10.1103/PhysRevLett.129.167402}
}

@article{PhysRevLett.130.166903,
  title     = {Characterizing Anomalous High-Harmonic Generation in Solids},
  author    = {Yue, Lun and Gaarde, Mette B.},
  journal   = {Phys. Rev. Lett.},
  volume    = {130},
  issue     = {16},
  pages     = {166903},
  numpages  = {7},
  year      = {2023},
  month     = {Apr},
  publisher = {American Physical Society},
  doi       = {10.1103/PhysRevLett.130.166903},
  url       = {https://link.aps.org/doi/10.1103/PhysRevLett.130.166903}
}

@article{PhysRevLett.93.233002,
  title     = {Coulomb Asymmetry in Above-Threshold Ionization},
  author    = {Goreslavski, S. P. and Paulus, G. G. and Popruzhenko, S. V. and Shvetsov-Shilovski, N. I.},
  journal   = {Phys. Rev. Lett.},
  volume    = {93},
  issue     = {23},
  pages     = {233002},
  numpages  = {4},
  year      = {2004},
  month     = {Nov},
  publisher = {American Physical Society},
  doi       = {10.1103/PhysRevLett.93.233002},
  url       = {https://link.aps.org/doi/10.1103/PhysRevLett.93.233002}
}

@article{PhysRevLett.98.013901,
  title     = {Scaling of Wave-Packet Dynamics in an Intense Midinfrared Field},
  author    = {Tate, J. and Auguste, T. and Muller, H. G. and Sali\`eres, P. and Agostini, P. and DiMauro, L. F.},
  journal   = {Phys. Rev. Lett.},
  volume    = {98},
  issue     = {1},
  pages     = {013901},
  numpages  = {4},
  year      = {2007},
  month     = {Jan},
  publisher = {American Physical Society},
  doi       = {10.1103/PhysRevLett.98.013901},
  url       = {https://link.aps.org/doi/10.1103/PhysRevLett.98.013901}
}

@article{PhysRevLett.99.253903,
  title     = {Quantum Path Interference in the Wavelength Dependence of High-Harmonic Generation},
  author    = {Schiessl, K. and Ishikawa, K. L. and Persson, E. and Burgd\"orfer, J.},
  journal   = {Phys. Rev. Lett.},
  volume    = {99},
  issue     = {25},
  pages     = {253903},
  numpages  = {4},
  year      = {2007},
  month     = {Dec},
  publisher = {American Physical Society},
  doi       = {10.1103/PhysRevLett.99.253903},
  url       = {https://link.aps.org/doi/10.1103/PhysRevLett.99.253903}
}

@article{Weissenbilder2022,
  author  = {Weissenbilder, R.
             and Carlstr{\"o}m, S.
             and Rego, L.
             and Guo, C.
             and Heyl, C. M.
             and Smorenburg, P.
             and Constant, E.
             and Arnold, C. L.
             and L'Huillier, A.},
  title   = {How to optimize high-order harmonic generation in gases},
  journal = {Nat. Rev. Phys.},
  year    = {2022},
  month   = {Nov},
  day     = {01},
  volume  = {4},
  number  = {11},
  pages   = {713-722},
  issn    = {2522-5820},
  doi     = {10.1038/s42254-022-00522-7},
  url     = {https://doi.org/10.1038/s42254-022-00522-7}
}

@article{Xiao:22,
  author    = {ZhiLei Xiao and Wei Quan and ShaoGang Yu and XuanYang Lai and XiaoJun Liu and ZhengRong Wei and Jing Chen},
  journal   = {Opt. Express},
  number    = {9},
  pages     = {14873--14885},
  publisher = {Optica Publishing Group},
  title     = {Nonadiabatic strong field ionization of noble gas atoms in elliptically polarized laser pulses},
  volume    = {30},
  month     = {Apr},
  year      = {2022},
  url       = {https://opg.optica.org/oe/abstract.cfm?URI=oe-30-9-14873},
  doi       = {10.1364/OE.454846}
}

@article{Xie_2022,
  doi       = {10.1088/1361-6455/ac8033},
  url       = {https://dx.doi.org/10.1088/1361-6455/ac8033},
  year      = {2022},
  month     = {sep},
  publisher = {IOP Publishing},
  volume    = {55},
  number    = {18},
  pages     = {185002},
  author    = {Xuejiao Xie and Ruihua Xu and Fabin Zhang and Shujuan Yu and Xun Liu and Wei Li and Yanjun Chen},
  title     = {Coulomb effects on time-trajectory-resolved high-order harmonic generation},
  journal   = {J. Phys. B}
}

@article{Xie:20,
  author    = {Xuejiao Xie and Chao Chen and Guoguo Xin and Jie Liu and Yanjun Chen},
  journal   = {Opt. Express},
  number    = {22},
  pages     = {33228--33239},
  publisher = {Optica Publishing Group},
  title     = {Coulomb-induced ionization time lag after electrons tunnel out of a barrier},
  volume    = {28},
  month     = {Oct},
  year      = {2020},
  url       = {https://opg.optica.org/oe/abstract.cfm?URI=oe-28-22-33228},
  doi       = {10.1364/OE.408424}
}

@article{Xie:25,
  author    = {Xuejiao Xie and Shujuan Yu and Yanpeng Li and Chunyan Zhang and Zengqiang Yang and Yanjun Chen},
  journal   = {Opt. Express},
  number    = {7},
  pages     = {14702--14715},
  publisher = {Optica Publishing Group},
  title     = {Odd-even harmonic emission from oriented NO molecule with nodal planes},
  volume    = {33},
  month     = {Apr},
  year      = {2025},
  url       = {https://opg.optica.org/oe/abstract.cfm?URI=oe-33-7-14702},
  doi       = {10.1364/OE.555856}
}

@article{PhysRevLett.105.253002,
  title     = {Low-Energy Structures in Strong Field Ionization Revealed by Quantum Orbits},
  author    = {Yan, Tian-Min and Popruzhenko, S. V. and Vrakking, M. J. J. and Bauer, D.},
  journal   = {Phys. Rev. Lett.},
  volume    = {105},
  issue     = {25},
  pages     = {253002},
  numpages  = {4},
  year      = {2010},
  month     = {Dec},
  publisher = {American Physical Society},
  doi       = {10.1103/PhysRevLett.105.253002},
  url       = {https://link.aps.org/doi/10.1103/PhysRevLett.105.253002}
}

@article{PhysRevLett.63.2212,
  title     = {Tunneling ionization of noble gases in a high-intensity laser field},
  author    = {Augst, S. and Strickland, D. and Meyerhofer, D. D. and Chin, S. L. and Eberly, J. H.},
  journal   = {Phys. Rev. Lett.},
  volume    = {63},
  issue     = {20},
  pages     = {2212--2215},
  numpages  = {0},
  year      = {1989},
  month     = {Nov},
  publisher = {American Physical Society},
  doi       = {10.1103/PhysRevLett.63.2212},
  url       = {https://link.aps.org/doi/10.1103/PhysRevLett.63.2212}
}

@article{PhysRevLett.109.083002,
  title     = {Probing the Longitudinal Momentum Spread of the Electron Wave Packet at the Tunnel Exit},
  author    = {Pfeiffer, A. N. and Cirelli, C. and Landsman, A. S. and Smolarski, M. and Dimitrovski, D. and Madsen, L. B. and Keller, U.},
  journal   = {Phys. Rev. Lett.},
  volume    = {109},
  issue     = {8},
  pages     = {083002},
  numpages  = {5},
  year      = {2012},
  month     = {Aug},
  publisher = {American Physical Society},
  doi       = {10.1103/PhysRevLett.109.083002},
  url       = {https://link.aps.org/doi/10.1103/PhysRevLett.109.083002}
}

@article{PhysRevLett.116.063003,
  title     = {Ionization Time and Exit Momentum in Strong-Field Tunnel Ionization},
  author    = {Teeny, Nicolas and Yakaboylu, Enderalp and Bauke, Heiko and Keitel, Christoph H.},
  journal   = {Phys. Rev. Lett.},
  volume    = {116},
  issue     = {6},
  pages     = {063003},
  numpages  = {5},
  year      = {2016},
  month     = {Feb},
  publisher = {American Physical Society},
  doi       = {10.1103/PhysRevLett.116.063003},
  url       = {https://link.aps.org/doi/10.1103/PhysRevLett.116.063003}
}

@article{PhysRevLett.127.273201,
  title     = {Nonadiabatic Strong Field Ionization of Atomic Hydrogen},
  author    = {Trabert, D. and Anders, N. and Brennecke, S. and Sch\"offler, M. S. and Jahnke, T. and Schmidt, L. Ph. H. and Kunitski, M. and Lein, M. and D\"orner, R. and Eckart, S.},
  journal   = {Phys. Rev. Lett.},
  volume    = {127},
  issue     = {27},
  pages     = {273201},
  numpages  = {6},
  year      = {2021},
  month     = {Dec},
  publisher = {American Physical Society},
  doi       = {10.1103/PhysRevLett.127.273201},
  url       = {https://link.aps.org/doi/10.1103/PhysRevLett.127.273201}
}

@article{10.3389fphy.2023.1120654,
  author  = {Wang, Zhiqiang  and Quan, Wei  and Hao, Xiaolei  and Chen, Jing  and Liu, Xiaojun },
  title   = {The ellipticity dependence of Rydberg state excitation of noble gas atoms subject to strong laser fields},
  journal = {Front. Phys.},
  volume  = {11},
  pages   = {1120654},
  year    = {2023},
  url     = {https://www.frontiersin.org/journals/physics/articles/10.3389/fphy.2023.1120654},
  issn    = {2296-424X}
}

@article{PhysRevLett.105.133002,
  title     = {Direct Test of Laser Tunneling with Electron Momentum Imaging},
  author    = {Arissian, L. and Smeenk, C. and Turner, F. and Trallero, C. and Sokolov, A. V. and Villeneuve, D. M. and Staudte, A. and Corkum, P. B.},
  journal   = {Phys. Rev. Lett.},
  volume    = {105},
  issue     = {13},
  pages     = {133002},
  numpages  = {4},
  year      = {2010},
  month     = {Sep},
  publisher = {American Physical Society},
  doi       = {10.1103/PhysRevLett.105.133002},
  url       = {https://link.aps.org/doi/10.1103/PhysRevLett.105.133002}
}

@article{Boroumand_2022,
  doi       = {10.1088/1361-6455/ac9205},
  url       = {https://dx.doi.org/10.1088/1361-6455/ac9205},
  year      = {2022},
  month     = {oct},
  publisher = {IOP Publishing},
  volume    = {55},
  number    = {21},
  pages     = {213001},
  author    = {Boroumand, N and Thorpe, A and Parks, A M and Brabec, T},
  title     = {Keldysh ionization theory of atoms: mathematical details},
  journal   = {J. Phys. B}
}

@article{2004Nonadiabatic,
  title   = {Nonadiabatic quantum path analysis of high-order harmonic generation: Role of the carrier-envelope phase on short and long paths},
  author  = { Sansone, G.  and  Vozzi, C.  and  Stagira, S.  and  Nisoli, M. },
  journal = {Phys. Rev. A},
  volume  = {70},
  number  = {1},
  pages   = {013411},
  year    = {2004}
}

@article{2013Wavelength,
  title   = {Wavelength Scaling of High Harmonic Generation Close to the Multiphoton Ionization Regime},
  author  = { Lai, Chien Jen  and  Cirmi, Giovanni  and  Hong, Kyung Han  and  Moses, Jeffrey  and  Huang, Shu Wei  and  Granados, Eduardo  and  Keathley, Phillip  and  Bhardwaj, Siddharth  and  Kaertner, Franz X. },
  journal = {Phys. Rev. Lett.},
  number  = {1/8},
  volume  = {111},
  pages   = {073901},
  year    = {2013}
}

@article{2012Strong,
  title   = {Strong-field approximation for the wavelength scaling of high-harmonic generation},
  author  = { Austin, Dane R.  and  Biegert, Jens },
  journal = {Phys. Rev. A},
  volume  = {86},
  number  = {2},
  pages   = {023813},
  year    = {2012}
}

@article{2013High,
  title   = {High-order harmonic generation from field-distorted orbitals},
  author  = { Spiewanowski, Maciej Dominik  and  Etches, Adam  and  Madsen, Lars Bojer },
  journal = {Phys. Rev. A},
  volume  = {87},
  number  = {4},
  pages   = {043424},
  year    = {2013}
}

@article{2023Revealing,
  title   = {Revealing the nonadiabatic tunneling dynamics in solid-state high harmonic generation},
  author  = { Zuo, R.  and  Song, Xiaohong  and  Ben, S.  and  Meier, T.  and  Yang, Weifeng },
  journal = {Phys. Rev. Research},
  year    = {2023},
  volume  = {5},
  pages   = {L022040}
}

@article{Auguste_2012,
  doi       = {10.1088/1367-2630/14/10/103014},
  url       = {https://dx.doi.org/10.1088/1367-2630/14/10/103014},
  year      = {2012},
  month     = {oct},
  publisher = {IOP Publishing},
  volume    = {14},
  number    = {10},
  pages     = {103014},
  author    = {Auguste, T and Catoire, F and Agostini, P and DiMauro, L F and Chirila, C C and Yakovlev, V S and Salières, P},
  title     = {Driving-frequency scaling of high-harmonic quantum paths},
  journal   = {New J. Phys.}
}

@article{Ghimire2019,
  author   = {Ghimire, Shambhu
              and Reis, David A.},
  title    = {High-harmonic generation from solids},
  journal  = {Nature Physics},
  year     = {2019},
  month    = {Jan},
  day      = {01},
  volume   = {15},
  number   = {1},
  pages    = {10-16},
  issn     = {1745-2481},
  doi      = {10.1038/s41567-018-0315-5},
  url      = {https://doi.org/10.1038/s41567-018-0315-5}
}

@article{Luu2018,
  author   = {Luu, Tran Trung
              and Yin, Zhong
              and Jain, Arohi
              and Gaumnitz, Thomas
              and Pertot, Yoann
              and Ma, Jun
              and W{\"o}rner, Hans Jakob},
  title    = {Extreme--ultraviolet high--harmonic generation in liquids},
  journal  = {Nature Communications},
  year     = {2018},
  month    = {Sep},
  day      = {13},
  volume   = {9},
  number   = {1},
  pages    = {3723},
  issn     = {2041-1723},
  doi      = {10.1038/s41467-018-06040-4},
  url      = {https://doi.org/10.1038/s41467-018-06040-4}
}

@article{Mondal2023,
  author   = {Mondal, Angana
              and Neufeld, Ofer
              and Yin, Zhong
              and Nourbakhsh, Zahra
              and Svoboda, V{\'i}t
              and Rubio, Angel
              and Tancogne-Dejean, Nicolas
              and W{\"o}rner, Hans Jakob},
  title    = {High-harmonic spectroscopy of low-energy electron-scattering dynamics in liquids},
  journal  = {Nature Physics},
  year     = {2023},
  month    = {Dec},
  day      = {01},
  volume   = {19},
  number   = {12},
  pages    = {1813-1820},
  issn     = {1745-2481},
  doi      = {10.1038/s41567-023-02214-0},
  url      = {https://doi.org/10.1038/s41567-023-02214-0}
}

@article{doi:10.1126/science.1163439,
  author  = {P. Eckle  and A. N. Pfeiffer  and C. Cirelli  and A. Staudte  and R. Dörner  and H. G. Muller  and M. Büttiker  and U. Keller },
  title   = {Attosecond Ionization and Tunneling Delay Time Measurements in Helium},
  journal = {Science},
  volume  = {322},
  number  = {5907},
  pages   = {1525-1529},
  year    = {2008},
  doi     = {10.1126/science.1163439},
  url     = {https://www.science.org/doi/abs/10.1126/science.1163439}
}

@article{PhysRevA.95.053425,
  title     = {Experimental verification of the nonadiabatic effect in strong-field ionization with elliptical polarization},
  author    = {Li, Min and Liu, Ming-Ming and Geng, Ji-Wei and Han, Meng and Sun, Xufei and Shao, Yun and Deng, Yongkai and Wu, Chengyin and Peng, Liang-You and Gong, Qihuang and Liu, Yunquan},
  journal   = {Phys. Rev. A},
  volume    = {95},
  issue     = {5},
  pages     = {053425},
  numpages  = {8},
  year      = {2017},
  month     = {May},
  publisher = {American Physical Society},
  doi       = {10.1103/PhysRevA.95.053425},
  url       = {https://link.aps.org/doi/10.1103/PhysRevA.95.053425}
}

@article{ZAIR2013184,
  title    = {Molecular internal dynamics studied by quantum path interferences in high order harmonic generation},
  journal  = {Chemical Physics},
  volume   = {414},
  pages    = {184-191},
  year     = {2013},
  note     = {Attosecond spectroscopy},
  issn     = {0301-0104},
  doi      = {https://doi.org/10.1016/j.chemphys.2012.12.022},
  url      = {https://www.sciencedirect.com/science/article/pii/S0301010412004739},
  author   = {Amelle Zaïr and Thomas Siegel and Suren Sukiasyan and Francois Risoud and Leonardo Brugnera and Christopher Hutchison and Zsolt Diveki and Thierry Auguste and John W.G. Tisch and Pascal. Salières and Misha Y. Ivanov and Jonathan P. Marangos}
}

@article{PhysRevA.61.063801,
  title     = {Nonadiabatic three-dimensional model of high-order harmonic generation in the few-optical-cycle regime},
  author    = {Priori, E. and Cerullo, G. and Nisoli, M. and Stagira, S. and De Silvestri, S. and Villoresi, P. and Poletto, L. and Ceccherini, P. and Altucci, C. and Bruzzese, R. and de Lisio, C.},
  journal   = {Phys. Rev. A},
  volume    = {61},
  issue     = {6},
  pages     = {063801},
  numpages  = {8},
  year      = {2000},
  month     = {May},
  publisher = {American Physical Society},
  doi       = {10.1103/PhysRevA.61.063801},
  url       = {https://link.aps.org/doi/10.1103/PhysRevA.61.063801}
}

@article{Li2025,
  author  = {Li, Mingxuan
             and Tang, Xiangyu
             and Wang, Huiyong
             and Li, Jialong
             and Wang, Wentao
             and Cai, Jiaao
             and Zhang, Jieda
             and San, Xinyue
             and Zhao, Xinning
             and Ma, Pan
             and Luo, Sizuo
             and Jin, Cheng
             and Ding, Dajun},
  title   = {Efficient generation of Bessel-Gauss attosecond pulse trains via nonadiabatic phase-matched high-order harmonics},
  journal = {Light: Science {\&} Applications},
  year    = {2025},
  month   = {May},
  day     = {06},
  volume  = {14},
  number  = {1},
  pages   = {181},
  issn    = {2047-7538},
  doi     = {10.1038/s41377-025-01845-7},
  url     = {https://doi.org/10.1038/s41377-025-01845-7}
}

@article{PhysRevA.80.013401,
  title     = {Quantitative rescattering theory for high-order harmonic generation from molecules},
  author    = {Le, Anh-Thu and Lucchese, R. R. and Tonzani, S. and Morishita, T. and Lin, C. D.},
  journal   = {Phys. Rev. A},
  volume    = {80},
  issue     = {1},
  pages     = {013401},
  numpages  = {23},
  year      = {2009},
  month     = {Jul},
  publisher = {American Physical Society},
  doi       = {10.1103/PhysRevA.80.013401},
  url       = {https://link.aps.org/doi/10.1103/PhysRevA.80.013401}
}

@article{PhysRevA.87.023418,
  title     = {Above-threshold ionization with highly charged ions in superstrong laser fields. II. Relativistic Coulomb-corrected strong-field approximation},
  author    = {Klaiber, Michael and Yakaboylu, Enderalp and Hatsagortsyan, Karen Z.},
  journal   = {Phys. Rev. A},
  volume    = {87},
  issue     = {2},
  pages     = {023418},
  numpages  = {11},
  year      = {2013},
  month     = {Feb},
  publisher = {American Physical Society},
  doi       = {10.1103/PhysRevA.87.023418},
  url       = {https://link.aps.org/doi/10.1103/PhysRevA.87.023418}
}

@inbook{Yan2013,
  author    = {Yan, T.-M.
               and Popruzhenko, S. V.
               and Bauer, D.},
  editor    = {Yamanouchi, Kaoru
               and Midorikawa, Katsumi},
  title     = {Trajectory-Based Coulomb-Corrected Strong Field Approximation},
  booktitle = {Progress in Ultrafast Intense Laser Science: Volume IX},
  year      = {2013},
  publisher = {Springer Berlin Heidelberg},
  address   = {Berlin, Heidelberg},
  pages     = {1--16},
  isbn      = {978-3-642-35052-8},
  doi       = {10.1007/978-3-642-35052-8_1},
  url       = {https://doi.org/10.1007/978-3-642-35052-8_1}
}

@article{PhysRevA.78.023814,
  title     = {Extraction of the species-dependent dipole amplitude and phase from high-order harmonic spectra in rare-gas atoms},
  author    = {Le, Anh-Thu and Morishita, Toru and Lin, C. D.},
  journal   = {Phys. Rev. A},
  volume    = {78},
  issue     = {2},
  pages     = {023814},
  numpages  = {6},
  year      = {2008},
  month     = {Aug},
  publisher = {American Physical Society},
  doi       = {10.1103/PhysRevA.78.023814},
  url       = {https://link.aps.org/doi/10.1103/PhysRevA.78.023814}
}

@article{PhysRevA.91.023428,
  title     = {Quantum path interference in the wavelength-dependent below-threshold harmonic generation},
  author    = {He, Lixin and Lan, Pengfei and Zhai, Chunyang and Li, Yang and Wang, Zhe and Zhang, Qingbin and Lu, Peixiang},
  journal   = {Phys. Rev. A},
  volume    = {91},
  issue     = {2},
  pages     = {023428},
  numpages  = {6},
  year      = {2015},
  month     = {Feb},
  publisher = {American Physical Society},
  doi       = {10.1103/PhysRevA.91.023428},
  url       = {https://link.aps.org/doi/10.1103/PhysRevA.91.023428}
}

@article{PhysRevLett.112.113002,
  title     = {Classical-Quantum Correspondence for Above-Threshold Ionization},
  author    = {Li, Min and Geng, Ji-Wei and Liu, Hong and Deng, Yongkai and Wu, Chengyin and Peng, Liang-You and Gong, Qihuang and Liu, Yunquan},
  journal   = {Phys. Rev. Lett.},
  volume    = {112},
  issue     = {11},
  pages     = {113002},
  numpages  = {5},
  year      = {2014},
  month     = {Mar},
  publisher = {American Physical Society},
  doi       = {10.1103/PhysRevLett.112.113002},
  url       = {https://link.aps.org/doi/10.1103/PhysRevLett.112.113002}
}

@article{PhysRevA.94.013415,
  title     = {Semiclassical two-step model for strong-field ionization},
  author    = {Shvetsov-Shilovski, N. I. and Lein, M. and Madsen, L. B. and R\"as\"anen, E. and Lemell, C. and Burgd\"orfer, J. and Arb\'o, D. G. and T\ifmmode \mbox{\H{o}}\else \H{o}\fi{}k\'esi, K.},
  journal   = {Phys. Rev. A},
  volume    = {94},
  issue     = {1},
  pages     = {013415},
  numpages  = {12},
  year      = {2016},
  month     = {Jul},
  publisher = {American Physical Society},
  doi       = {10.1103/PhysRevA.94.013415},
  url       = {https://link.aps.org/doi/10.1103/PhysRevA.94.013415}
}

@article{PhysRevA.92.043407,
  title     = {Influence of the Coulomb potential on above-threshold ionization: A quantum-orbit analysis beyond the strong-field approximation},
  author    = {Lai, X.-Y. and Poli, C. and Schomerus, H. and Faria, C. Figueira de Morisson},
  journal   = {Phys. Rev. A},
  volume    = {92},
  issue     = {4},
  pages     = {043407},
  numpages  = {12},
  year      = {2015},
  month     = {Oct},
  publisher = {American Physical Society},
  doi       = {10.1103/PhysRevA.92.043407},
  url       = {https://link.aps.org/doi/10.1103/PhysRevA.92.043407}
}

@article{PhysRevA.65.011804,
  title     = {Dramatic extension of the high-order harmonic cutoff by using a long-wavelength driving field},
  author    = {Shan, Bing and Chang, Zenghu},
  journal   = {Phys. Rev. A},
  volume    = {65},
  issue     = {1},
  pages     = {011804},
  numpages  = {4},
  year      = {2001},
  month     = {Dec},
  publisher = {American Physical Society},
  doi       = {10.1103/PhysRevA.65.011804},
  url       = {https://link.aps.org/doi/10.1103/PhysRevA.65.011804}
}

@article{PhysRevLett.103.073902,
  title     = {Wavelength Scaling of High Harmonic Generation Efficiency},
  author    = {Shiner, A. D. and Trallero-Herrero, C. and Kajumba, N. and Bandulet, H.-C. and Comtois, D. and L\'egar\'e, F. and Gigu\`ere, M. and Kieffer, J-C. and Corkum, P. B. and Villeneuve, D. M.},
  journal   = {Phys. Rev. Lett.},
  volume    = {103},
  issue     = {7},
  pages     = {073902},
  numpages  = {4},
  year      = {2009},
  month     = {Aug},
  publisher = {American Physical Society},
  doi       = {10.1103/PhysRevLett.103.073902},
  url       = {https://link.aps.org/doi/10.1103/PhysRevLett.103.073902}
}

@article{Li_2016,
  doi       = {10.1088/0953-4075/49/7/075603},
  url       = {https://doi.org/10.1088/0953-4075/49/7/075603},
  year      = {2016},
  month     = {mar},
  publisher = {IOP Publishing},
  volume    = {49},
  number    = {7},
  pages     = {075603},
  author    = {Li, Y P and Yu, S J and Duan, X Y and Shi, Y Z and Chen, Y J},
  title     = {Wavelength dependence of high-harmonic yield from aligned molecules: roles of structure and electron dynamics},
  journal   = {J. Phys. B}
}

@article{PhysRevA.51.1495,
  title     = {Rings in above-threshold ionization: A quasiclassical analysis},
  author    = {Lewenstein, M. and Kulander, K. C. and Schafer, K. J. and Bucksbaum, P. H.},
  journal   = {Phys. Rev. A},
  volume    = {51},
  issue     = {2},
  pages     = {1495--1507},
  numpages  = {0},
  year      = {1995},
  month     = {Feb},
  publisher = {American Physical Society},
  doi       = {10.1103/PhysRevA.51.1495},
  url       = {https://link.aps.org/doi/10.1103/PhysRevA.51.1495}
}

@article{PhysRevA.107.043109,
  title     = {Roles of laser ellipticity in attoclocks},
  author    = {Che, J. Y. and Huang, J. Y. and Zhang, F. B. and Chen, C. and Xin, G. G. and Chen, Y. J.},
  journal   = {Phys. Rev. A},
  volume    = {107},
  issue     = {4},
  pages     = {043109},
  numpages  = {8},
  year      = {2023},
  month     = {Apr},
  publisher = {American Physical Society},
  doi       = {10.1103/PhysRevA.107.043109},
  url       = {https://link.aps.org/doi/10.1103/PhysRevA.107.043109}
}

@article{che2023,
  title   = {Advances in response time of strong-field ionization of atoms},
  author  = {Che, J. Y. and Chen, C. and Li, W. Y. and Li, W. and Chen, Y. J.},
  journal = {Acta Phys. Sin.},
  volume  = {72},
  issue   = {19},
  pages   = {193301},
  year    = {2023},
  doi     = {10.7498/aps.72.20230983},
  url     = {https://wulixb.iphy.ac.cn/cn/article/doi/10.7498/aps.72.20230983}
}

@misc{chen2022responsetimephotoemissionquantumclassic,
  title         = {Response time of photoemission at quantum-classic boundary},
  author        = {C. Chen and J. Y. Che and W. Y. Li and S. Wang and X. J. Xie and J. Y. Huang and Y. G. Peng and G. G. Xin and Y. J. Chen},
  year          = {2022},
  eprint        = {2111.08491},
  archiveprefix = {arXiv},
  primaryclass  = {physics.atom-ph},
  url           = {https://arxiv.org/abs/2111.08491}
}

@article{PhysRevA.111.053118,
  title     = {Coulomb-related symmetry in laser-induced tunneling ionization of atoms and molecules},
  author    = {Chen, Z. Y. and Shen, S. Q. and Li, Y. P. and Yang, Z. Q. and Che, J. Y. and Chen, Y. J.},
  journal   = {Phys. Rev. A},
  volume    = {111},
  issue     = {5},
  pages     = {053118},
  numpages  = {15},
  year      = {2025},
  month     = {May},
  publisher = {American Physical Society},
  doi       = {10.1103/PhysRevA.111.053118},
  url       = {https://link.aps.org/doi/10.1103/PhysRevA.111.053118}
}

@dataset{sun_2026_18288498,
  author    = {Sun, Fengjian and
               Pi, Liang-Wen},
  title     = {Nonadiabatic effect in high-order harmonic
               generation revealed by a fully analytical method
               data
               },
  month     = jan,
  year      = 2026,
  publisher = {Zenodo},
  doi       = {10.5281/zenodo.18288498},
  url       = {https://doi.org/10.5281/zenodo.18288498}
}
\end{document}